\documentclass[a4paper,12pt]{article}
\pdfoutput=1
\usepackage{feynmp-auto,expdlist}
\usepackage{amsmath, amsfonts, amssymb}
\usepackage{graphicx}
\usepackage{enumerate}
\usepackage{latexsym}
\usepackage{hepnicenames}
\usepackage{enumerate}
\usepackage{soul}
\usepackage[normalem]{ulem}
\usepackage{bbold}
\usepackage{wasysym} 
\usepackage{makecell}
\RequirePackage[ colorlinks=true,urlcolor=blue,anchorcolor=blue,citecolor=blue,filecolor=blue,
               linkcolor=blue,menucolor=blue,linktocpage=true,pdfproducer=medialab,pagebackref=true]{hyperref}
 \hypersetup{
 colorlinks   = true, %Colours links instead of ugly boxes
 urlcolor     = PineGreen, %Colour for external hyperlinks
 linkcolor    = blue, %Colour of internal links
  citecolor   = blue %Colour of citations     
}

\newcommand{\blue}[1]{\color{blue}#1\color{black}}

\font\tenrsfs=rsfs10 at 12pt
\font\sevenrsfs=rsfs7
\font\fiversfs=rsfs5
\newfam\rsfsfam
\textfont\rsfsfam=\tenrsfs
\scriptfont\rsfsfam=\sevenrsfs
\scriptscriptfont\rsfsfam=\fiversfs

\numberwithin{equation}{section}

\usepackage{mathrsfs}
\usepackage{braket}
\usepackage{titling}
\usepackage{amsmath}
\usepackage{slashed}
\usepackage{amssymb}
\usepackage{epsfig}
\usepackage{graphicx}
\usepackage{color}
\usepackage{rotating}
\usepackage[margin=1.in]{geometry} 
\usepackage[table,xcdraw,dvipsnames]{xcolor}
\usepackage[compress,numbers,sort]{natbib}
\usepackage{colortbl}
\usepackage{pdflscape}
\usepackage{color}
\usepackage{mathtools}
\usepackage{colortbl}
\definecolor{Gray}{gray}{0.95}
\definecolor{RGray}{gray}{0.85}
\definecolor{CGray}{gray}{0.92}

\newcommand{\SU}{{\rm SU}}

\newcommand{\U}{{\rm U}}

\newcommand{\V}{{\cal V}}

\newcommand{\N}{{\cal N}}

\definecolor{nicered}{rgb}{0.7,0.1,0.1}
\definecolor{nicegreen}{rgb}{0.1,0.5,0.1}
\definecolor{red}{rgb}{1.0, 0, 0}
\definecolor{niceblue}{rgb}{0,0,0.8}
\definecolor{red}{rgb}{1.0, 0, 0}
\allowdisplaybreaks

\definecolor{rosso}{cmyk}{0,1,1,0.4}
\definecolor{rossos}{cmyk}{0,1,1,0.55}
\definecolor{rossoc}{cmyk}{0,1,1,0.2}
\definecolor{blu}{cmyk}{1,1,0,0.3}
\definecolor{blus}{cmyk}{1,1,0,0.6}
\definecolor{bluc}{cmyk}{1,1,0,0.1}
\definecolor{verde}{cmyk}{0.92,0,0.59,0.25}
\definecolor{verdec}{cmyk}{0.92,0,0.59,0.15}
\definecolor{verdes}{cmyk}{0.92,0,0.59,0.4}

\definecolor{orangeQCD}{rgb}{0.9725490196078431,0.6274509803921569,0.49411764705882355}
\definecolor{colZN}{rgb}{0.968735,0.465621,0.212103}
\definecolor{greenAstro}{rgb}{0.000000,0.427451,0.172549}
\definecolor{greenAstroL}{rgb}{0.0, 0.66, 0.42}
\definecolor{blueEDM}{rgb}{0.031373,0.317647,0.611765}
\definecolor{BlueC0}{rgb}{0.000000,0.447059,0.698039}
\definecolor{GreenC1}{rgb}{0.000000,0.619608,0.450980}
\definecolor{OrangeC2}{rgb}{0.835294,0.368627,0.000000}
\def\eq#1{{Eq.~(\ref{#1})}}
\def\eqs#1#2{{Eqs.~(\ref{#1})--(\ref{#2})}}
\def\fig#1{{Fig.~\ref{#1}}}

\def\sect#1{{Sect.~\ref{#1}}}

\def\vev#1{\left\langle #1\right\rangle}
\def\abs#1{\left| #1\right|}

\def\det{\mbox{det}\,}

\renewcommand{\bar}{\overline}

\newcommand{\beq}{\begin{equation}}
\newcommand{\eeq}{\end{equation}}
\newcommand{\bea}{\begin{eqnarray}}
\newcommand{\eea}{\end{eqnarray}}

\renewcommand{\[}{\left[}
\renewcommand{\]}{\right]}
\renewcommand{\(}{\left(}
\renewcommand{\)}{\right)}

\newcommand{\Q}{\mathcal{Q}}

\renewcommand{\S}{\mathcal{S}}

\usepackage[capitalise]{cleveref}
\usepackage{nicefrac}
\usepackage{subcaption}\small 
\newcommand{\email}[1]{\href{mailto:#1}{\tt #1}}

\begin{document}
\vspace{1.5cm}

{\flushright
{\blue{ \hfill}\\
\blue{ \hfill}\\
\vspace{-1.cm}
\blue{DESY 21-010}\\
\blue{IFT-UAM/CSIC-20-143}\\
\blue{FTUAM-20-21}}\\
\hfill 
}

\begin{center}
{\Large\LARGE\Huge\bf\color{blus} 
An even lighter QCD axion}\\[1cm]

{\bf Luca Di Luzio$^{a}$, Belen Gavela$^{b,\,c}$, Pablo Quilez$^{a}$, Andreas Ringwald$^{a}$}\\[7mm]

{\it $^a$Deutsches Elektronen-Synchrotron DESY, \\ 
Notkestra\ss e 85, D-22607 Hamburg, Germany}\\[1mm]
{\it $^b$Departamento de F\'isica Te\'orica, Universidad Aut\'onoma de Madrid, \\ 
Cantoblanco, 28049, Madrid, Spain}\\[1mm]
{\it $^c$Instituto de F\'isica Te\'orica, IFT-UAM/CSIC, \\
Cantoblanco, 28049, Madrid, Spain}\\[1mm]

\vspace{0.5cm}
\begin{abstract}
We explore whether the axion which solves the strong CP problem can naturally 
be much lighter than the canonical QCD axion. The  $Z_\N$ symmetry proposed 
by Hook, with $\N$ mirror and degenerate worlds coexisting in Nature and linked 
by the axion field, is considered in terms of generic effective axion couplings. 
We show that  the total potential is safely approximated  by a single cosine 
in the large $\N$ limit, and we determine the analytical formula for the exponentially 
suppressed axion mass. The resulting universal enhancement of all axion interactions 
relative to those of the canonical QCD axion has a strong impact on the prospects of 
axion-like particle experiments such as  ALPS II, IAXO and many others:  experiments searching for generic axion-like particles  have in fact  discovery potential to solve the strong CP problem. The finite density axion potential is  also analyzed 
and we show that the $Z_\N$ asymmetric background of high-density stellar 
environments sets already significant model-independent constraints: 
$3\le\N\lesssim47$ for an axion scale  $f_a\lesssim 2.4\times10^{15}$ GeV, with tantalizing 
discovery prospects for any value of $f_a$ and down to $\N\sim9$ 
with future neutron star and gravitational wave data,  down to the ultra-light mass region. 
In addition, two specific ultraviolet $Z_\N$ completions are developed: 
a composite axion one and a KSVZ-like model with improved Peccei-Quinn quality.

\end{abstract}

\begin{minipage}[l]{.9\textwidth}
{\footnotesize \vspace{1cm}
\begin{center}
\textit{E-mail:} 
\email{luca.diluzio@desy.de}\,,
\email{belen.gavela@uam.es}\,, \\
\email{pablo.quilez@desy.de}\,,
\email{andreas.ringwald@desy.de}
\end{center}}
\end{minipage} 
\thispagestyle{empty}
\bigskip

\end{center}

\setcounter{footnote}{0}

\newpage
\tableofcontents

\newpage

\section{Introduction}
\label{sec:intro}

The axion experimental program is 
in a blooming phase, 
with several new experiments and detection concepts promising the exploration 
of regions of parameter space  thought to be 
unreachable until a decade ago. 
Many of those experiments are 
simply prototypes,
awaiting the jump to become `big-experiments', 
or, in the case of more consolidated techniques, 
they are still far from saturating their full physics potential.  
Nonetheless, they sometimes reach sensitivities 
which go well-beyond astrophysical limits, 
albeit often still far from the customary  QCD axion window. 

On the other hand, since axion couplings are inherently ultraviolet (UV) dependent, 
such early stage experiments already provide valuable probes of the 
QCD axion parameter space.  
 Imagine for definiteness that ALPS II 
would detect a signal in 2021, 
would it be possible to interpret that
 as an axion that solves  the strong CP problem?
 Since the strong CP problem is 
one of the strongest 
motivations 
for 
new physics, if an axion-like particle (ALP)  will be 
ever discovered, there or elsewhere, it would be 
 compelling to explore whether it had something
to do with the strong CP problem. 
This work explores whether 
 wide regions in the ALP parameter space, well outside the traditional QCD axion band, may correspond to solutions of the strong CP problem. This is a question of profound theoretical and experimental relevance.

In axion solutions to the strong CP problem\footnote{That is, via a global 
chiral $\U(1)$ symmetry, exact although hidden (aka spontaneously broken) at the classical level and explicitly broken by instanton effects at the quantum level~\cite{Peccei:1977hh,Peccei1977}.} both the axion mass 
and  the couplings to ordinary matter scale as $1/f_a$, where $f_a$ is the 
axion decay constant, denoting the scale of new physics. 
The precise relation between mass and decay constant
depends on the characteristics of the strong interacting sector of the theory. When QCD is the only confining group to which the axion $a$ couples, in which case we denote the axion mass as $m_a^{\rm QCD}$, 
they are necessarily linked by the relation~\cite{Weinberg:1977ma,Wilczek1978}
\begin{equation}
m_a^{\rm QCD}  = \frac{\sqrt{\chi_{\rm QCD}}}{f_a} \simeq m_\pi \, f_\pi\,\frac{\sqrt{m_u\,m_d}}{m_u+m_d}\,\frac{1}{f_a }  \,,
\label{invisiblesaxion}
\end{equation}
where $\chi_{\rm QCD}, m_\pi, f_\pi, m_u$ and $m_d$ denote respectively the QCD topological susceptibility, the pion mass, its decay constant, and the up  and down quark masses.  Equation~(\ref{invisiblesaxion})  is completely model-independent as far as QCD is the only source of the axion mass, and  it defines the ``canonical QCD axion'', also often called ``invisible axion''.
 For  this 
  axion the   $a G_{\mu\nu}\tilde G^{\mu\nu}$ coupling to the gluon strength $G_{\mu\nu}$
 is directly responsible for the axion mass, 
 since the only source of explicit breaking
 of the global axial Peccei-Quinn (PQ) symmetry $\U(1)_{\rm PQ}$ is its QCD anomaly. The strength of other axion couplings to Standard Model (SM) fields is instead model-dependent: it varies with the matter content of the UV complete axion model. 
 
\begin{figure}[ht]
\centering
\includegraphics[width=15cm]{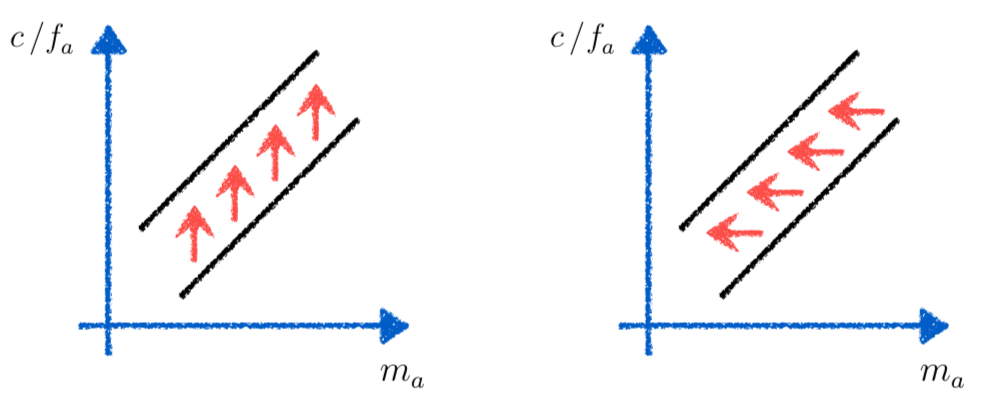} 
\caption{\small Different approaches to enlarge the parameter space of axions that solve the strong CP problem.  The canonical QCD axion relation is represented by the lower 
black line in the $\{ m_a, c/f_a\}$ parameter space, where $c$ denotes a generic effective axion coupling. 
Vertical displacements, possible within pure QCD axion models (i.e.~$m_a=m_a^{\rm QCD}$), are depicted on the left. Horizontal displacements (via enlarged strong gauge sectors) are illustrated on the right for the case of a 
lighter than usual axion to be explored here.
}
\label{fig:couplvsmass-new}       
\end{figure}

 In recent years there have been many attempts to enlarge the canonical QCD 
axion window, by considering UV  completions of the axion effective Lagrangian 
which departed from the minimal DFSZ \cite{Zhitnitsky:1980tq,Dine:1981rt} and KSVZ \cite{Kim:1979if,Shifman:1979if} constructions. Most approaches actually focussed on the possibility of modifying 
the Wilson coefficient of  specific axion-SM effective operators~\cite{DiLuzio:2016sbl,Farina:2016tgd,DiLuzio:2017pfr,Agrawal:2017cmd,Marques-Tavares:2018cwm,DiLuzio:2020wdo,Darme:2020gyx}.
That is,  the size of the coupling coefficients, at fixed $f_a$, is modified. 
 This has for example allowed 
 to populate new regions of the parameter space by moving vertically the axion band in 
 the  axion mass versus coupling plane, see \fig{fig:couplvsmass-new} left.
The results are  then ``channel specific'': different couplings $c$ are modified differently.  

The  parameter space of solutions can be alternatively changed 
 by  varying the axion mass at fixed $f_a$. This corresponds to horizontal displacements of the canonical axion band in the parameter space, see right panel in \fig{fig:couplvsmass-new}. It always requires that the magnitude  of the relation between the axion mass $m_a$ and $1/f_a$  departs from that in Eq.~(\ref{invisiblesaxion}): the confining sector of the SM must be enlarged beyond QCD.  New instanton sources  give then additional contributions 
 to the right-hand side of \eq{invisiblesaxion}.
  The practical consequence is a universal modification of the parameter space of {\it all} axion couplings at a given $m_a$, at variance with the vertical displacement scenarios.  This feature could {\it a priori} allow for the two mechanisms in \fig{fig:couplvsmass-new} to be distinguished.\footnote{For instance, via the measurement of the axion coupling to 
  the neutron electric dipole moment (nEDM) 
  operator at CASPER-electric \cite{Budker:2013hfa,JacksonKimball:2017elr}, in case the axion would also account for dark matter (DM).
  % (see Ref.~\cite{ZNDMpaper}). 
  The axion-to-nEDM coupling directly follows from the $m_a$--$f_a$ relation and 
  so it is unmodified in standard approaches to axion coupling enhancements 
  (left panel in \fig{fig:couplvsmass-new})
  that still rely on \eq{invisiblesaxion}.}

Examples of horizontal enlargement of the parameter space towards the right of the canonical QCD axion band are heavy axion models that solve the strong CP problem 
at low scales (e.g.~$f_a \sim$ TeV)~\cite{rubakov:1997vp,Berezhiani:2000gh,Gianfagna:2004je,Hsu:2004mf,Hook:2014cda,Fukuda:2015ana,Chiang:2016eav,Dimopoulos:2016lvn,Gherghetta:2016fhp,Kobakhidze:2016rwh,Agrawal:2017ksf,Agrawal:2017evu,Gaillard:2018xgk,Buen-Abad:2019uoc,Hook:2019qoh,Csaki:2019vte,Gherghetta:2020ofz}.
The present work explores  instead left horizontal shifts:
 true axions that solve the strong CP problem with $m_a \ll m_a^{\rm QCD}$.
 This avenue is more challenging, 
 since it requires a new source of PQ breaking aligned with QCD,
  whose contribution to the axion mass needs to \emph{almost} cancel that from QCD  without relying on fine-tunings. 

A possible mechanism to achieve this lighter-than-usual true axion in a technically natural way was recently 
put forth by Hook \cite{Hook:2018jle}, 
in terms of a discrete $Z_\N$ symmetry. $\N$ mirror and degenerate worlds would coexist in Nature, linked by an 
axion field\footnote{This setup for $\N = 2$ had previously led instead to an {\it enhancement} of the axion mass~\cite{Giannotti:2005eb}, because the axion field was assumed to be invariant under the $Z_2$ 
transformation.} which implements non-linearly the $Z_\N$ symmetry. One of those worlds is  our SM  one.
All the confining sectors contribute now to the right-hand side of Eq.~(\ref{invisiblesaxion}), conspiring by symmetry to suppress the axion mass without spoiling the solution to the strong CP problem. 
 The direct consequence  
is, for fixed $f_a$, a $\N$-dependent reduced axion mass   
 in spite of all confining scales  being equal to $\Lambda_{\rm QCD}$. 
  In other words, for a given value of $m_a$ it follows a universal enhancement of all axion interactions relative to those of the canonical QCD axion.  In this paper, we expand on
  the mathematical properties of the  implementation of the $Z_\N$ symmetry and  determine the analytic form of the exponential suppression of the axion mass and its potential in the large $\N$ limit. The phenomenological analysis of the number of possible mirror worlds $\N$ will be next carried 
out with present and projected data.  
  
 The study will also explore the $Z_\N$ axion potential at finite density, to confront present constraints and prospects from very dense stellar objects and gravitational waves.  It has been recently pointed out in 
 \cite{Hook:2017psm,Huang:2018pbu} that a generic reduced-mass axion  leads to strong effects on those systems, raising the effective mass in the dense media. In the scenario considered here, a stellar background made only of SM matter is by nature $Z_\N$-asymmetric: we will show analytically how such an asymmetric background breaks the cancellations which guaranteed an exponentially suppressed axion mass for the $Z_\N$ symmetric vacuum potential. 
Limits on the number of possible worlds will be obtained in turn.

 The theoretical framework to be used throughout the  work described above
is that 
 of effective axion couplings. Nevertheless, two  concrete UV completions of the $Z_\N$ scenario under consideration will be developed as well: a model  {\`a la KSVZ}~\cite{Kim:1979if,Shifman:1994ee}, and a composite model {\`a la Choi-Kim}~\cite{Kim:1984pt,Choi:1985cb}.  
The status of the Peccei-Quinn (PQ) quality problem will be also addressed. 

An important remark is that we will consider in this paper experiments that can test the solution to the strong CP problem without further assumptions.  Indeed, it is most relevant to get a clear panorama on the strong CP problem by itself, given its fundamental character. In particular, we {\it will not} discuss axion or ALP experiments that {\it do} rely on the assumption that the DM of the Universe may be constituted by axions.  
The cosmological evolution of the axion field in the $Z_\N$ scenario under discussion 
and its contribution to the DM relic abundance
departs drastically from the standard case,  
and it is discussed in 
a companion 
paper~\cite{ZNDMpaper}. 

The structure of the present paper can be easily inferred from the Table of Contents.

\section{Down-tuning the axion mass} 
\label{sec:downtuningma}

 In Ref.~\cite{Hook:2018jle} it was shown
 how to naturally down-tune the axion mass from its natural QCD value  in  Eq.~(\ref{invisiblesaxion}), 
exploiting the analyticity structure of the QCD axion potential in the presence of a $Z_\N$ symmetry.
For pedagogical purposes, before turning to the generic $Z_\N$ case we  analyze the (unsuccessful) case of a $Z_2$ symmetry: the SM plus one degenerate mirror world linked by an axion which realizes the symmetry non-linearly.

\subsection{The $Z_2$ case} 
\label{sec:Z2potential}

\begin{figure}[ht]
\centering
\includegraphics[width=0.75\textwidth]{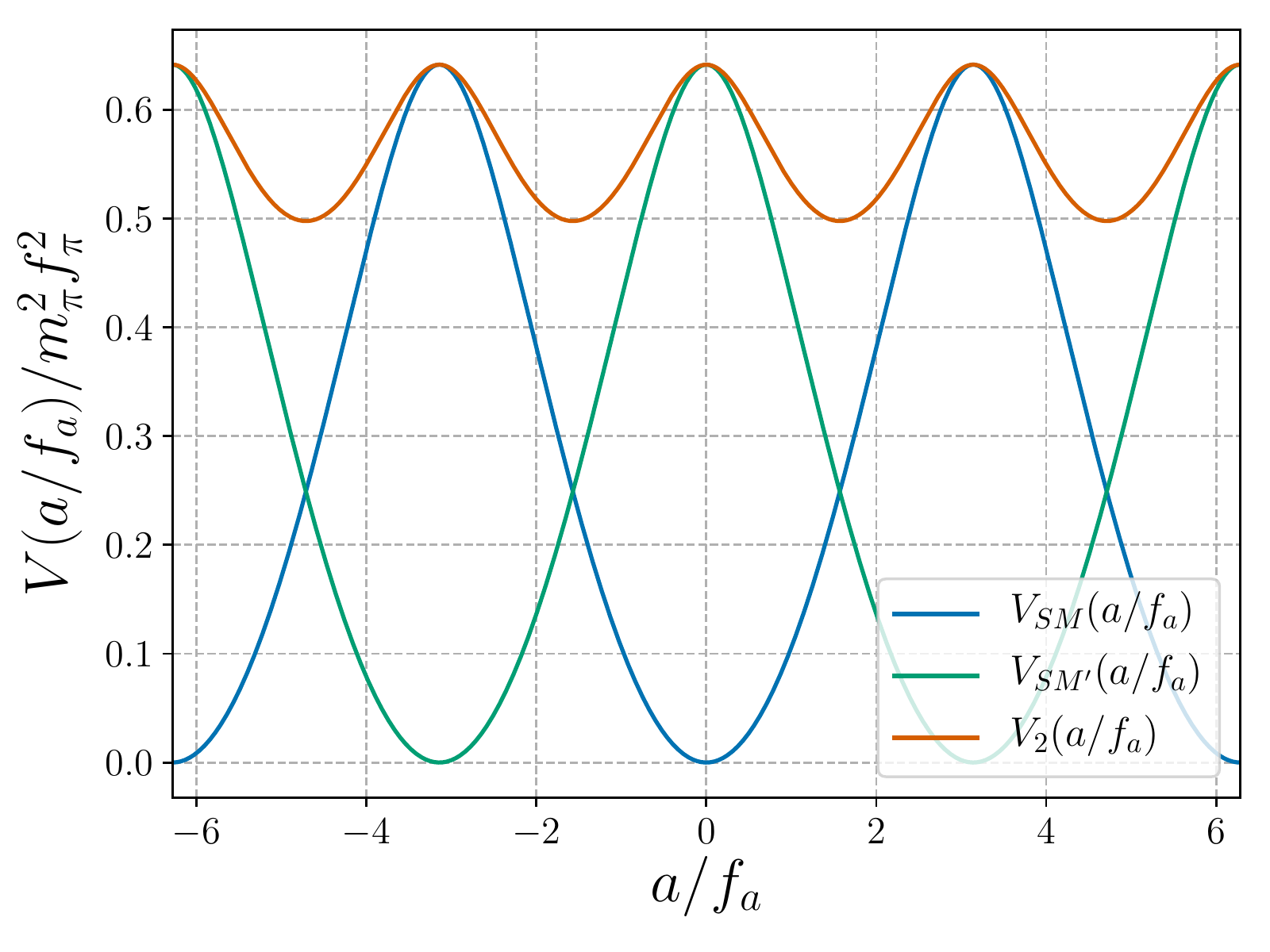} 
\caption{\small $Z_2$ axion potential. The mirror contribution to the axion potential $V_{\text{SM}'}(a/f_a)$ (in {\color{GreenC1} \bf green}) partially cancels that of the SM, $V_{\text{SM}}(a/f_a)$ (in  {\color{BlueC0} \bf blue}), leading to a total shallower potential $V_2(a/f_a)$ (in {\color{OrangeC2} \bf orange}). The total potential has a maximum in $a/f_a = 0$ and thus this $Z_2$ axion does not solve the SM strong CP problem. }
\label{fig:typicalaxionpotZ2}       
\end{figure}
Consider the SM plus a complete copy SM$'$, related  via a $Z_2$ symmetry which exchanges each SM field with its mirror counterpart, while the axion field is shifted by $\pi$:
\begin{align}
	Z_2:\quad &\text{SM}\longleftrightarrow \text{SM}' \\
	     & a \longrightarrow a+\pi f_a\,.
\end{align}
The Lagrangian, including the anomalous effective couplings of the axion to SM fields, then reads
\begin{equation}
	\mathcal{L}=\mathcal{L}_{\rm{SM}}+\mathcal{L}_{\rm{SM'}}+  \frac{\alpha_s}{8\pi}\,\Big(\frac{a}{f_{a}}-\theta \Big) G\widetilde{G}+  \frac{\alpha_s}{8\pi}\,\Big(\frac{a}{f_{a}}-\theta +\pi \Big) G'\widetilde{G}' + \dots \,,
	\label{Eq:LagZ2}
\end{equation}
where $\theta$ parametrizes the  anomalous QCD coupling, $\alpha_s$ is the QCD fine-structure constant, the Lorentz indices of the field strength $G_{\mu\nu}$ have been obviated, and the dots stand for possible $Z_2$-symmetric portals between the two mirror worlds 
(see \sect{sec:SMportals}). 
Without loss of generality, we can perform a uniform shift in $a$ such that the $\theta$ term in 
\eq{Eq:LagZ2} is set to zero. Therefore, the effective $\theta$-parameter of the SM 
corresponds to  
$\theta_\text{eff} \equiv \langle a \rangle /f_a$, where $\langle a \rangle$ denotes the vacuum expectation value (vev) of the axion field.

In the case of an exact $Z_2$ symmetry, 
all  couplings and masses of the mirror world and the SM would coincide  with the exception of the effective $\theta$-parameter. 
It is this difference (namely the $\pi$ shift in the effective $\theta$-parameters of the SM and its mirror) the one responsible for displaced contributions to the total axion potential, with destructive interference effects. Were the QCD axion potential to be a simple cosine, the total potential would vanish because the two contributions  (from QCD and mirror QCD)  would have exactly the same size but opposite sign, i.e.~$\propto \text{cos}({a}/{f_{a}})$ and  $\propto \text{cos}({a}/{f_{a}} +\pi)=-\text{cos}({a}/{f_{a}})$ respectively. However, for the 
true chiral axion potential~\cite{DiVecchia:1980yfw,Leutwyler:1992yt,diCortona:2015ldu}  the exact cancellation disappears and a residual potential --and thus a non-zero axion mass-- remains, 
which at leading chiral order reads (keeping only two flavours)
\begin{equation}
		V_2(a) =-\frac{m_{\pi}^{2} f_{\pi}^{2}}{m_u+m_d}\left\{\sqrt{m_{u}^{2}+m_{d}^{2}+2 m_{u} m_{d} \cos \left(\frac{a}{f_{a}}\right)} +\sqrt{m_{u}^{2}+m_{d}^{2}-2 m_{u} m_{d} \cos \left(\frac{a}{f_{a}} \right)}\right\}\,.
\end{equation}		
This $Z_2$-symmetric world would not solve the strong CP problem, though, because 
$a/f_a = 0$
is a maximum of the axion potential, as illustrated in Fig.~\ref{fig:typicalaxionpotZ2}. Indeed, as already pointed out in Ref.~\cite{Hook:2018jle}, 
$a/f_a = 0$ is a minimum of the potential 
only for odd values of $\N$, while it is a maximum for $\N$. Thefore, the simplest viable axion model that solves the strong CP problem with a reduced axion mass incorporates a $Z_3$ symmetry.

\subsection{$Z_\N$ axion} 
\label{sec:ZNpotential}

We consider now $\N$ copies of the SM that are interchanged under 
a $Z_\N$ symmetry which is non-linearly realized by the axion field: 
	\begin{align}
	Z_\N:\quad &\text{SM}_{k} \longrightarrow \text{SM}_{k+1\,(\text{mod} \,\N)} \label{mirror-charges}\\
	     & a \longrightarrow a + \frac{2\pi k}{\N} f_a\,,
	     \label{axion-detuned-charge}
\end{align}
with  $ k=0,\ldots, \N-1$.  One of those worlds will be our SM one. 
The most general Lagrangian implementing this symmetry describes $\N$ mirror worlds whose couplings take exactly the same values as in the SM, with the exception of the 
effective $\theta$-parameter: 
 for each copy  the effective $\theta$  value is shifted by $2\pi/\N$ with respect to that in the 
neighbour $k$ sector,
\begin{equation}
\label{Eq: Lagrangian ZN}
\mathcal{L} 
= 
\sum_{k=0}^{\N-1} \[ \mathcal{L}_{\text{SM}_k} +
\frac{\alpha_s}{8\pi}  \( \theta_a + \frac{2 \pi k}{\N} \) G_k \widetilde G_k \] \,+ \dots 
\end{equation}
where $\mathcal{L}_{\text{SM}_k}$ denotes  exact  copies of the SM total Lagrangian excluding the strong anomalous coupling, and  the dots stand for  $Z_\N$-symmetric portal 
couplings that may connect those different sectors 
(to be discussed in \sect{sec:SMportals}). 
In this equation  
$\theta_a \equiv a / f_a$ is the angular axion field 
defined in the interval $[-\pi,\pi)$, 
and  a universal (equal for all $k$ sectors) 
bare theta parameter has been set to zero
via an overall shift of the axion field.   
The SM is identified from now on with the $k=0$ sector: to ease the notation, 
 the label $k=0$ on SM quantities will be often dropped below.  
 Each QCD$_k$ sector contributes  to the $\theta_a$ potential, 
which in the 2-flavour leading order chiral expansion reads
\beq 
\label{eq:VNpot}
V_\N(\theta_a) = - A \sum_{k=0}^{\N-1} \sqrt{1 + z^2 + 2 z \cos{\( \theta_a + \frac{2\pi k}{\N}\)}} \, ,
\eeq
where
\beq
z \equiv m_u / m_d \approx 0.48\,, \qquad A \equiv \Sigma_0 m_d \approx \chi_0 (1+z) / z\,,
\eeq
and
\beq
\label{condensate}
\Sigma_{0}\equiv - \vev{\bar u u} = - \vev{\bar d d} =m_{\pi}^{2} f_{\pi}^{2}/(m_u+m_d)
\eeq
denotes the chiral condensate~\cite{Leutwyler:1992yt}, while  $\chi_0 \approx (75 \  \text{MeV})^4$ is the zero temperature QCD 
topological susceptibility \cite{Borsanyi:2016ksw,diCortona:2015ldu}.
Alternatively, the total $Z_\N$ axion potential can be written as
\begin{align}
V_\N(\theta_a)=-m_{\pi}^{2} f_{\pi}^{2}\sum_{k=0}^{\N-1}  \sqrt{1-\beta \sin ^{2}\left(\frac{\theta_a}{2}+\frac{\pi k}{\N}\right)}\, ,
\label{Eq:Vsmilga ZN}
\end{align}
where $\beta\equiv4m_um_d/(m_u+m_d)^2= 4z/(1+z)^2$.
 
\begin{figure}[ht]
\centering
\includegraphics[width=0.75\textwidth]{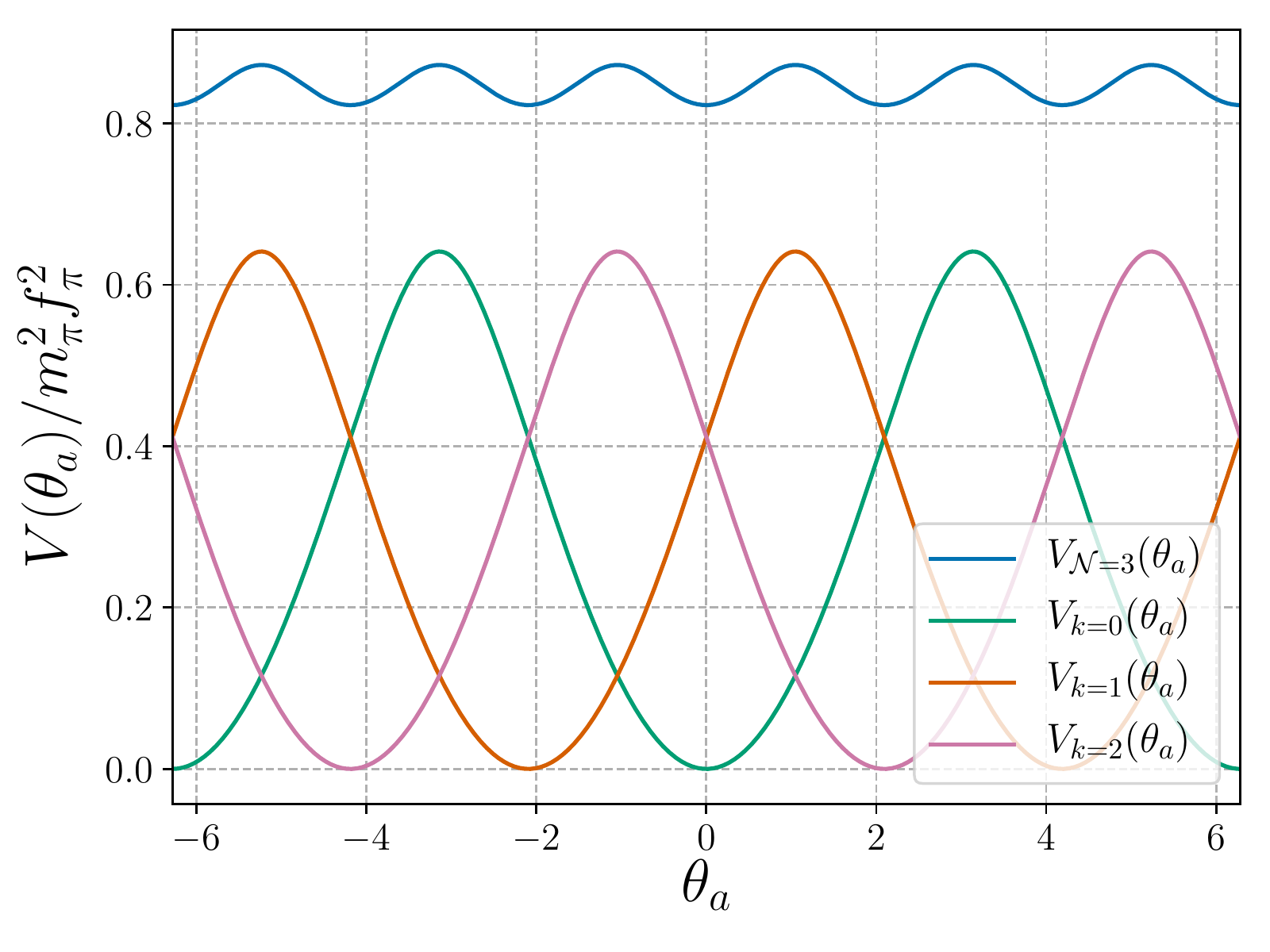} 
\caption{\small $Z_3$ axion potential. The contributions from the $\N=3$ worlds partially cancel each other, leading to an exponentially small total potential $V_{\N=3}(\theta_a)$ (in  {\color{BlueC0} \bf blue}) 
that exhibits a minimum in $\theta_a=0$.} 
\label{fig:typicalaxionpot}       
\end{figure}

For any $\N$,  $\theta_a=0$ is  an extrema of the axion potential. Indeed, using the property 
$\sin ({2 \pi(\N-k)}/{\N})=-\sin \left({2\pi k}/{\N}\right)$ 
it is straightforward to see that 
\begin{equation}
\left.\frac{\partial V_\N(\theta_a)}{\partial \theta_a}\right|_{\theta_a=0}=\frac{m_{\pi}^{2} f_{\pi}^{2}}{f_{a}} \frac{\beta}{4} \sum_{k=0}^{\N-1} \frac{\sin \left(\frac{2 \pi k}{\N}\right)}{\sqrt{1-\beta \sin ^{2}\left(\frac{\pi k}{\N}\right)}}=0\,.
\end{equation}
 The same holds for any $\theta_a= 2\pi n/\N $ with $n \in \mathbb{Z}$, because of the periodicity of the potential. 
  For $\N$ odd the potential $V(\theta_a)$ has $\N$ 
minima  
located at 
 \begin{equation}
 \theta_a =\{\pm {2\pi \ell/\N}\}   \qquad \qquad \text{for} \quad \ell=0,1,\dots,\frac{\N-1}{2}\,,
 \end{equation}
which includes the origin $\theta_a=0$,  
while for $\N$ even the origin becomes a maximum. 
This result --valid for any $\N$-- can be shown for instance using the exact Fourier series expansion of the potential in Eqs.~(\ref{eq:VNpot})-(\ref{Eq:Vsmilga ZN}) (see final part of \cref{App: Hypergeometric complete mass dependence}).
It follows that $\N$ odd is required  in order to solve the SM strong CP problem 
(albeit with a $1/\N$ tuning in the cosmological evolution \cite{Hook:2018jle,ZNDMpaper}). 
 The $k \neq 0$ 
worlds  have instead non-zero effective $\theta$-parameters: $\theta_k \equiv 2\pi k/\N$ for
$\vev{\theta_a} = 0$, see \eq{Eq: Lagrangian ZN}.    
 A typical shape of the axion potential for $\N = 3$ 
is illustrated in \fig{fig:typicalaxionpot}.  

 The different  effective $\theta_k$ values  translate into  slightly different  masses for  the pion mass in each mirror world, $m_{\pi} (\theta_k)$. At quadratic order in $m_\pi$ a reduction factor of up to $\sim \sqrt{3}$ 
results~\cite{diCortona:2015ldu},
 \begin{equation}
m_{\pi}^2 (\theta_k) = m_\pi^2\, \sqrt{1-\frac{4 m_{u} m_{d}}{\left(m_{u}+m_{d}\right)^{2}} \sin ^{2}\left(\frac{\pi k}{\mathcal{N}}\right)}\,.
\end{equation} 
 Interestingly, nuclear physics would 
be drastically different in the different mirror copies. In particular,  a new scalar pion ($\pi_k$)  to nucleon ($N_k$) coupling is generated in all worlds but the SM one (see e.g.~Refs.~\cite{Ubaldi:2008nf,Lee:2020tmi}):
\beq 
\label{eq:LchiPTNucl}
\mathcal{L}_{\chi\text{PT}} 
\supset c_+  \sum_{k=0}^{\N-1} \frac{m_u m_d \sin\theta_k}{[m_u^2 + m_d^2 + 2 m_u m_d \cos\theta_k]^{1/2}} 
\frac{\pi_k^a}{f_\pi} \bar N_k \tau^a N_k \, ,
\eeq
where $c_+$ is an $\mathcal{O}(1)$ low-energy constant 
of the baryon chiral Lagrangian. Its impact on the cosmological histories of the mirror worlds is discussed 
in Ref.~\cite{ZNDMpaper} for the $Z_\N$ scenario under discussion.

Overall, for our world to be that with vanishing effective $\theta$, the $\sim 10$ orders of magnitude tuning required by the SM strong CP problem
has been traded by a $1/\N$ adjustment,  while $\N$ could a priori be as low as $\N=3$.\footnote{Although we work in the exact $Z_\N$ limit, 
cosmological considerations require the temperature of the SM thermal bath  to be higher than that of the other $k\neq 0$ 
sectors~\cite{Berezhiani:2000gw,Berezhiani:2003xm,Foot:2014mia}.    
% from various cosmological considerations (e.g.~number of relativistic degrees of freedom). 
 Mechanisms to achieve these different temperatures 
%between the SM and the other 
%$k\neq 0$ 
%sectors 
 will be discussed in Ref.~\cite{ZNDMpaper}.}

\subsubsection{Renormalizable portals to the SM} 
\label{sec:SMportals}

Renormalizable portals between the SM and its mirror copies (left implicit in Eq.~(\ref{Eq: Lagrangian ZN})) 
are allowed by the $Z_\N$ symmetry. In the following, we classify for completeness the portal operators connecting the different $k$ 
sectors. 

\subsubsection*{Higgs portals}
The most general $Z_\N$ symmetric scalar potential for the Higgs doublets $H_k$  of the different mirror worlds includes terms of the form
\begin{align} 
\V(H_k) &\supset 
\sum_{i=1}^{(\N-1)/2} 
\kappa_i
\sum_{k=0}^{\N-1} 
\( \abs{H_k}^2 - \frac{v^2}{2} \)\( \abs{H_{k+i}}^2 - \frac{v^2}{2} \)\Big|_{\,(\text{mod} \,\N)} \, ,
\label{Eq: Higgs portals ZN}
\end{align}
where  $v $ denotes the Higgs vev and $\kappa_i$ are 
dimensionless parameters. Note that the $Z_\N$-symmetric mixings between different worlds may include next-neighbour, next-to-next neighbour etc.~interactions. 
 All $\kappa_ {i\ge 1}$ terms provide renormalizable portals 
between the mirror Higgs copies ($H_{k\neq 0}$) 
and the SM Higgs ($H_{k = 0}$).  

\subsubsection*{Kinetic mixing}
Terms mixing the $\U(1)^k_Y$ hypercharge field strengths of mirror worlds are {\it a priori }also allowed by the $Z_\N$ symmetry,  
\begin{equation}
	\mathcal{L}\supset 
 \sum_{i=1}^{(\N-1)/2}  
\epsilon_i
\sum_{k=0}^{\N-1}  F^{\mu\nu}_kF_{\mu\nu,\,k+i} \big|_{\,(\text{mod} \,\N)} \,,
\label{Eq: Kinetic portals ZN}
\end{equation}
where $F^{\mu\nu}_k$ denote here the $k$-hypercharge field strenghts  and $\epsilon_i$ are free dimensionless parameters.

The above renormalizable portals are subject to strong
cosmological constraints, as discussed in Ref.~\cite{ZNDMpaper}. 
This can suggest a `naturalness' issue for the Higgs and the kinetic portal couplings, as they cannot be forbidden in terms of 
internal symmetries. Nevertheless, such small couplings  may be technically natural because of an enhanced Poincar\'e symmetry \cite{Foot:2013hna,Volkas:1988cm}: 
in the limit where non-renormalizable interactions are neglected,  the $\kappa_{i\ne 0}$ and $\epsilon_{i\ne 0} \to 0$ 
limit corresponds to an enhanced $\mathcal{P}^\N$  
symmetry (namely an independent space-time Poincar\'e transformation 
$\mathcal{P}$
in each sector). Those couplings are then protected 
 from receiving radiative corrections other than those induced by 
the explicit $\mathcal{P}^\N$ breaking due to 
gravitational and axion-mediated interactions, which are presumably small.  In addition, other terms in the scalar potential which depend on the details of the UV completion of the $Z_\N$ axion scenario may be present and strongly constrained; an example is given below in Sect.~\ref{KSVZ}.

\subsection{Axion potential in the large $\N$ limit}

It is non-trivial to sum the series which defines the axion potential, 
 Eq.~(\ref{Eq:Vsmilga ZN}). However,  the presence of the $Z_\N$ symmetry allows for the application of powerful mathematical tools related to its Fourier decomposition and holomorphicity properties, that lead to simplified expressions in the large $\N$ limit. 

 \subsubsection{Holomorphicity bounds and 
 convergence of Riemannian sums}
 \label{Subsec: Holomorphicity bounds}
As  first noticed  in Ref.~\cite{Hook:2018jle}, the fact that the potential in \cref{Eq:Vsmilga ZN} corresponds to a  Riemann sum allows one to express it as an integral plus subleading terms,
\begin{equation}
	V_{\mathcal{N}}\left(\theta_{a}\right)= \sum_{k=0}^{\N-1} V\left(\theta_a+\frac{2 \pi k}{\mathcal{N}}\right) = \frac{\N}{2\pi}\int_0^{2\pi}V(x)dx + \mathcal{O}(\N^0)\,, 
	\label{Eq: riemanian sum}
\end{equation}
where the definition of each single-world potential, $V\left(\theta_a+\frac{2 \pi k}{\mathcal{N}}\right)$, 
can be read off 
Eq.~(\ref{eq:VNpot}). Most importantly, the integral does not depend on the field $\theta_a$ and the amplitude of the axion potential is thus solely contained in the subleading terms. The latter are nothing but the error $E$ committed in approximating the Riemann sum by an integral,
\begin{align}
E_{\N}(V)=\int_{0}^{2 \pi} V(x) d x-\frac{2 \pi}{\N} \sum_{k=0}^{\N-1} V\left(\theta_a+\frac{2 \pi k}{\N}\right)\,.
\end{align}
Powerful theorems exist that describe the fast convergence of this approximation.
 It can be shown, applying complex analysis,
  that if some conditions are satisfied the convergence of the rectangular rule is exponential (see e.g.~Section 3 in Ref.~\cite{doi:10.1137/130932132}). More precisely, if $V(\theta_a)$ is a $2\pi$-periodic function and 
 it can be extended to a holomorphic function $V(w)$  in a rectangle from 0 to $2\pi$ and from $-ib$ to $+ib$, then the error of the rectangular rule is constrained as  
 \begin{align}
\left|E_{\N}(V)\right| \leq \frac{4 \pi M}{e^{\N b}-1}\,, 
\label{Eq: theorem En}
 \end{align}
 where $M$ is an upper limit on $V(w)$ in the rectangular region defined above. 
 As a consequence, the axion mass will be exponentially suppressed for large $\N$. 
 More in detail,  let us apply  the theorem to the second derivative of the potential,
\begin{align}
V''(\theta_a)=-\frac{m_{\pi}^{2} f_{\pi}^{2}}{2} \frac{z}{1+z}  \frac{2\left(1+z^{2}\right) \cos \left(\theta_a\right)+z\left[3+\cos \left(\theta_a/2\right)\right]}{\left[1+z^{2}+2 z \cos \left(\theta_a\right)\right]^{3 / 2}}\,,
\label{Eq:V''}
\end{align}
which can be extended in the complex plane to a holomorphic function until the expression under the square root vanishes. Indeed, this function has branch points in\footnote{This result coincides with that in Ref.~\cite{Hook:2018jle}, which defines $a=\log (c+\sqrt{c^{2}-1})$, for $c=\frac{\left(m_{u}+m_{d}\right)^{2}}{2 m_{u} m_{d}}-1$. Note that the variable $a$ can be simplified as $a=\log (m_d/m_u)=-\log z$.}
\begin{align}
w_{cut}=\pi \pm i \log z\,.
\end{align}
Naively, it is tempting to apply the theorem assuming $b= \log z$ in Eq.~(\ref{Eq: theorem En}). This is not possible though,  since $V''(w)$ is not bounded in the rectangular 
region, due to a divergence in the branch point. As we show in \cref{app:math}, 
it is possible to optimize the bound obtained above on the axion mass ($V''(\theta_a)/f_a^2$) by allowing a departure from $\log z$, $b= \log z+ \Delta b$, which leads to 
\begin{align}
\Delta b= \frac{3}{2}\frac{1}{\N}\,, 
\end{align}
where the factor $3/2$ stems from the order of the divergence of \cref{Eq:V''} in the branch point $w_{cut}$. Implementing this result in \cref{Eq: theorem En}, it follows that
\begin{align}
m_a^2f^2_a \leq \left|E_{\N}(V'')\right| \leq \pi m_{\pi}^{2} f_{\pi}^{2} \sqrt{\frac{1-z}{1+z}}\left(\frac{2}{3}\right)^{3/2}\N^{3/2}\frac{1}{e^{-3/2}z^{-\N}-1}\,. 
\label{Eq: result theorem En}
 \end{align}
In \cref{fig:ma_numvsanalyt} we compare this analytical bound with the numerical result: our analytical bound captures the correct dependence on $\N$ of the $Z_\N$ axion mass,
\begin{equation}
	\label{Eq: holom axion mass}
	m_a^2f^2_a  \propto m_{\pi}^{2} f_{\pi}^{2} \,\sqrt{\frac{1-z}{1+z}} \,\N^{3/2}\, z^\N\,,
\end{equation}
although it misses the overall constant factor.
The overall factor will be analytically determined in the following. 
Nevertheless, the discussion above has the two-fold interest of determining the correct exponential suppression {\it and} of being very general, as it only relies on the holomorphicity structure of the potential, and not on the specific form it takes. As a consequence, {\it the exponential suppression of the 
axion mass is not spoiled 
when considering the subleading chiral corrections to \cref{Eq:Vsmilga ZN}}.

\subsubsection{Fourier expansion:  axion mass  from hypergeometric functions}
It is possible to gain further physical insight on the origin of the cancellations in the potential by constructing its Fourier series expansion. 
As  shown in \cref{App: fourier},  the Fourier series of any scalar potential respecting the $Z_\N$ shift symmetry only receives contributions from modes that are multiples of $\N$. Moreover, if the potential can be written as a sum of shifted contributions, as it is the case for the $Z_\N$ axion under discussion --see \cref{Eq: riemanian sum}-- then the Fourier series of the total potential $V_{\N}(\theta_a)$ can be easily obtained in terms of the Fourier series of a single $V(\theta_a)$ term,  
 leading to
\begin{align}
V_{\mathcal{N}}\left(\theta_{a}\right)=2 \N 
	\sum_{t=1}^{\infty} \hat V(t\N)  \cos (t\,\N\theta_a)\,,
\label{Eq: fourier potential general }
\end{align}
where $\hat V(n)$ denotes the coefficient of the Fourier series  for the single-world
potential $V(\theta_a)$,
\begin{align}
\hat V(n)=-  \frac{m_{\pi}^{2} f_{\pi}^{2}}{1+z}\int_0^{2\pi} \cos(n t)\sqrt{1+z^{2}+2 z \cos \left(t\right)}dt\,.
\label{Eq: fourier single world def}
\end{align}
 It is convenient to express this integral in terms of the Gauss hypergeometric function (see \cref{App: Hypergeometric complete mass dependence} 
 and Ref.~\cite{GRADSHTEYN1980904} for conventions and relevant properties),
\begin{align}
\hat V(n)=(-1)^{n+1}  \frac{m_{\pi}^{2} f_{\pi}^{2}}{1+z}\, z^{n} \,\frac{\Gamma(n-1/2)}{\Gamma(-1/2) \,n !}
\,_2F_{1}\left(\begin{array}{c}
-1/2,\, n-1/2\\
 n+1
\end{array} \bigg|z^2\right) \, .
\end{align}
As  shown in \cref{App: Hypergeometric complete mass dependence}, in the large $\N$ limit this expression  further simplifies to
\begin{align}
\hat V(n)\simeq\,(-1)^{n}\, \frac{m_{\pi}^{2} f_{\pi}^{2}}{2\sqrt{\pi}} \, \sqrt{\frac{1-z}{1+z}}n^{-3/2}\, z^{n}\,,
\end{align}
leading to the following expression for the total potential 
\begin{align}
V_{\mathcal{N}}\left(\theta_{a}\right) &\simeq  \frac{m_{\pi}^{2} f_{\pi}^{2}}{\sqrt{\pi}} \, \sqrt{\frac{1-z}{1+z}}\N^{-1/2}\, 
	\sum_{t=1}^{\infty} \,(-1)^{t\,\N}\, t^{-3/2}\, z^{t\,\N}  \cos (t\,\N\theta_a) \nonumber \\
	&\simeq \frac{m_{\pi}^{2} f_{\pi}^{2}}{\sqrt{\pi}} \, \sqrt{\frac{1-z}{1+z}}\N^{-1/2}\, 
	 \,(-1)^{\N}\,  z^{\N}  \cos (\N\theta_a)\,,
\label{Eq: fourier potential large N hyper}
\end{align}
where in the second line we have kept only the first mode in the expansion, 
as the higher modes are exponentially suppressed with respect to it.
 {\it The total potential is thus safely approximated by a single cosine}. 
It trivially follows from  Eq.~(\ref{Eq: fourier potential large N hyper}) 
that  $\theta_a=0$ is a minimum of the total potential for $\N$ odd, and a maximum for $\N$ even.   Here and all through this work purely constant terms in the potential are obviated, as they have no impact on the axion mass.

Eq.~(\ref{Eq: fourier potential large N hyper}) can be rewritten as 
  \beq 
  \label{Eq: fourier potential large N hyper-compact}
V_{\mathcal{N}}\left(\theta_a\right)
  \simeq - \frac{m_a^2 f_a^2}{\N^2} \,\cos (\N\theta_a)\,, 
\eeq 
where the $Z_\N$ axion mass $m_a$  in the large $\N$ limit is finally given by a compact and analytical formula, 
\begin{equation}
	\label{Eq: hyper axion mass}
	m_a^2 \, f^2_a  \simeq  \frac{m_{\pi}^{2} f_{\pi}^{2}}{\sqrt{\pi}} \,\sqrt{\frac{1-z}{1+z}} \,\N^{3/2} \,z^\N\,. 
\end{equation}  
The overall coefficient is thus determined, in addition to exhibiting the $z^\N$ exponential suppression of the potential and the specific $\N$ dependence previously argued in \cref{Eq: holom axion mass} from holomorphicity arguments.
In summary, in the large $\N$ limit the axion mass is reduced with respect to that 
of the QCD axion by a factor
\begin{figure}[ht]
\centering
\includegraphics[width=0.9\textwidth]{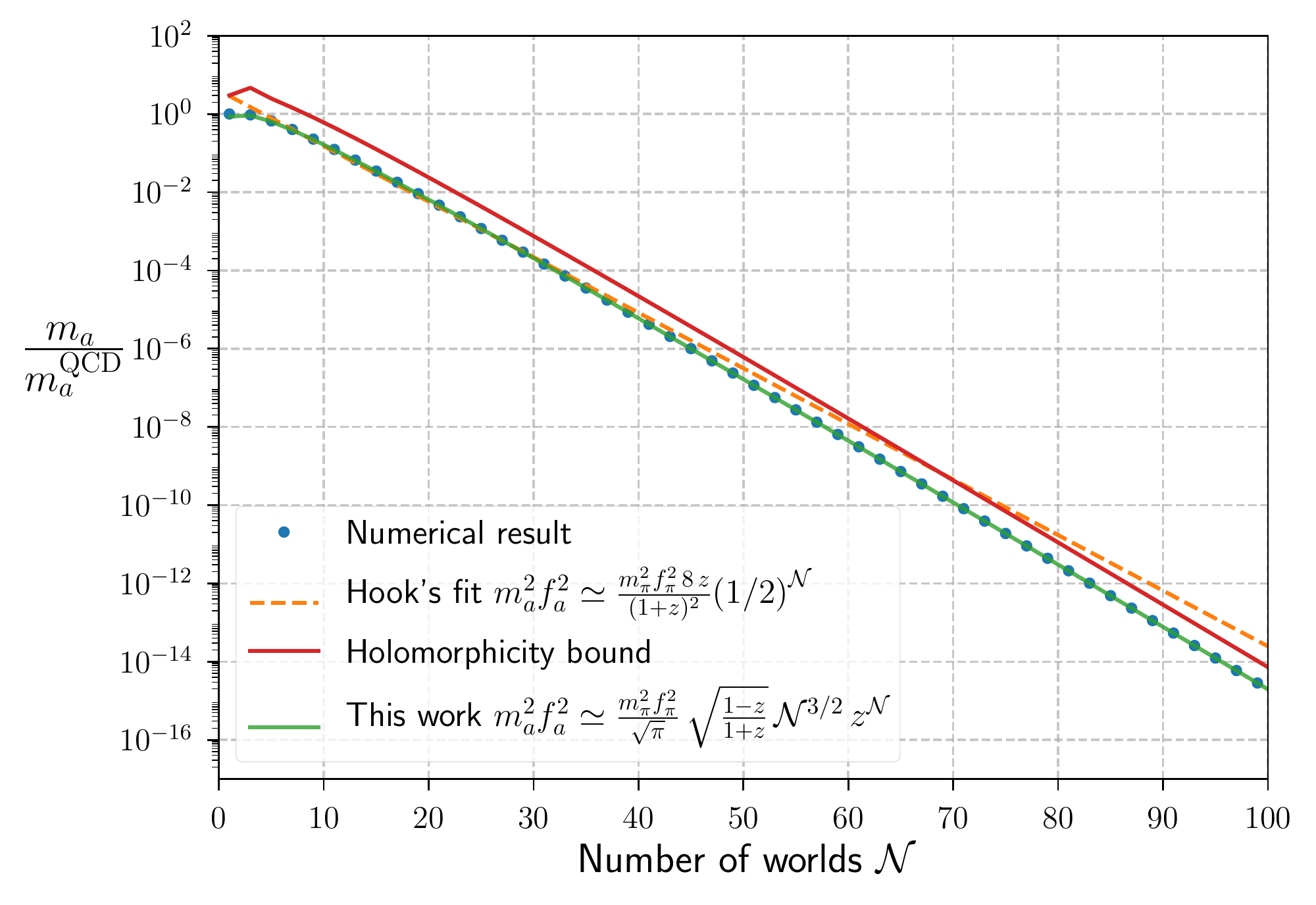} 
\caption{\small Comparison between different evaluations of the 
axion mass as a function of $\N$. Our large-$\N$ analytical result in \cref{eq:ratio-holo-mass} 
 ({\color{GreenC1} \bf green} curve) provides a remarkably good approximation to the numerical evaluation (dots).}
\label{fig:ma_numvsanalyt}       
\end{figure}
\begin{align}
\label{eq:ratio-holo-mass} 
\bigg(\frac{m_a}{m_a^{\rm QCD}}\bigg)^2 \simeq \frac{1}{\sqrt{\pi}}\sqrt{1-z^2} (1+z)\, \N^{3/2} z^{\N-1}\, \,,
\end{align}
where $m_a^{\rm QCD}$  
denotes the mass of the canonical QCD axion as given  in Eq.~(\ref{invisiblesaxion}).\footnote{Note that, although $\N=1$ denotes the SM world, $m_a^{\rm QCD}$ does not correspond to $\N=1$  in Eq.~(\ref{Eq: hyper axion mass}), because the latter  is only valid in the large $\N$ limit.} 
 This ratio is  illustrated in  \cref{fig:ma_numvsanalyt}, which compares the numerical behaviour with: $a)$ the analytical dependence previously proposed in Ref.~\cite{Hook:2018jle}; 
$b)$ that from the holomorphicity bound in Eq.~(\ref{Eq: result theorem En}); 
$c)$ the full analytical result in Eq.~(\ref{eq:ratio-holo-mass}).  Our analytical results improve on previous ones by Hook on a number of aspects: $i)$ the explicit determination of the exponential behavior controlled by $z^{\N}\sim 2^{-\N}$; $ii)$ the improved  $\N$ dependence from the factor $\N^{3/2}$; 
$iii)$ the $z$-dependence of the axion mass in $\sqrt{\frac{1-z}{1+z}}$; 
$iv)$ the determination of the prefactor  $1/\sqrt{\pi}$.

In practice, the large $\N$ results in Eqs.~(\ref{Eq: fourier potential large N hyper})-(\ref{eq:ratio-holo-mass}) turn out to be an excellent  approximation already for $\N=3$.

\section{UV completions and alternative scenarios} 
\label{sec:UVcompl}
 Up to this point, the analysis has been largely independent from the precise UV completion of the  $Z_\N$ axion scenario.  For the sake of illustration, in this section we provide  two 
 UV completions of the axion effective Lagrangian 
in \eq{Eq: Lagrangian ZN}. We also briefly discuss an alternative implementation of the $Z_\N$ symmetry in which the resulting axion is heavier than usual (rather than lighter).

\subsection{KSVZ $Z_\N$ axion}
\label{KSVZ}
Consider $\N$ copies of vector-like 
Dirac fermions 
$\Q_k$ ($k=0,\ldots,\N-1$) 
transforming in the fundamental representation of QCD$_k$, together 
with a gauge singlet complex scalar 
$\S$. 
The action of the $Z_\N$ symmetry on these fields is postulated to be  
\begin{align}
	Z_\N:\ \Q_k &\to \Q_{k+1\,(\text{mod} \,\N)} \, , \\
	     \quad \S &\to e^{2\pi i /\N}\S \, ,
	     \label{Eq:KSVZZNtransf}
\end{align}
while the SM Lagrangian and its copies obey Eq.~(\ref{mirror-charges}) under $Z_\N$.  
The most general Lagrangian containing the new degrees 
of freedom then reads 
\begin{equation}
	\mathcal{L}_{\rm UV} = 
	|\partial_\mu \S|^2 
	+ \sum_{k=0}^{\N-1} \[ \bar \Q_k i\slashed{\mathcal{D}} \Q_k
	+ y 
	e^{2\pi i k/\N} \S \bar \Q_k P_R \Q_k + \text{h.c.} \]
	- \V(\S,H_k) \, , 
	\label{Eq:LagUVZN}
\end{equation}
where $P_R \equiv (1+\gamma_5)/2$. It  exhibits an accidental $\U(1)_{\rm PQ}$ symmetry 
\begin{align}
	\U(1)_{\rm PQ}:\ \Q_k &\to e^{-i\gamma_5 \frac{\alpha}{2}}\Q_{k} \, , \\
	     \quad \S &\to e^{i \alpha}\S \, ,
	     \label{Eq:KSVZPQtransf}
\end{align}
that is spontaneously broken by the vev of $\S$, $v_\S$,  via a proper 
`mexican-hat' potential $\V(\S, H_k)$, whose structure is discussed below.  
Decomposing the $\S$ field in a polar basis, 
\beq 
\S = \frac{1}{\sqrt{2}} (v_\S + \rho) e^{i\frac{a}{v_\S}} \, , 
\eeq
in terms of canonically normalized radial  ($\rho$) and axion modes, 
the latter can be rotated away from the Yukawa term in \eq{Eq:LagUVZN} 
via an axion-dependent axial transformation 
\beq 
\label{eq:Qaxial}
\Q_k \to e^{-i\gamma_5 \( \frac{a}{2v_\S} + \frac{\pi k}{\N} \) }\Q_{k} \, .
\eeq
The heavy quarks, with real mass\footnote{Note that we crucially removed also 
the $k$-dependent phases from the Yukawas, in order to properly integrate out 
the heavy $\Q_k$ fields.} 
$m_{\Q_k} = \frac{y_\S v_\S}{\sqrt{2}}$, 
can next be integrated out in order to obtain the  low-energy axion effective field theory. 
Because the transformation in \eq{eq:Qaxial} is QCD$_k$ anomalous, 
with anomaly factor $2N_k = 1$, the resulting axion effective Lagrangian is given by 
\beq 
\label{eq:deltaLKSVZ}
\delta \mathcal{L}_{\rm UV} = 
\sum_{k=0}^{\N-1}
\frac{\alpha_s}{8\pi}   \( \frac{a}{v_\S} + \frac{2 \pi k}{\N} \) G_k \widetilde G_k \,,
\eeq
which yields precisely  \eq{Eq: Lagrangian ZN}, after the identification $v_\S = f_a$. 

Furthermore, the presence of 
the singlet scalar $\S$ introduces new scalar portals between the SM and its mirror worlds, in addition to the generic ones  in \cref{Eq: Higgs portals ZN}. The scalar potential in the latter equation
 should thus be enlarged by
\begin{equation}
\V(H_k) \longrightarrow  \V(\S, H_k) = \V(H_k) + \delta \V\,,
\end{equation}
 with 
\begin{align} 
\delta \V&= 
\lambda_{\S} \( \abs{\S}^2 - \frac{f_a^2}{2} \)^2 
+ \kappa_{\S} 
\( \abs{\S}^2 - \frac{f_a^2}{2} \) 
\sum_{k=0}^{\N-1} \( \abs{H_k}^2 - \frac{v^2}{2} \) \,.
\end{align}
Note that, because the Higgs vev $v$  is the same in all $k$ sectors 
 due to the unbroken $Z_\N$ symmetry,  the required  
hierarchy of scales is obtained with a single fine-tuning between $v$ and $f_a$, as for elementary canonical QCD axions.

It is also possible to choose the representations of the $\Q_k$ fields to 
transform non-trivially under the  electroweak$_k$ gauge groups, so that they could e.g.~mix 
with SM$_k$ quarks in a $Z_\N$ invariant way 
and decay efficiently in the early Universe, 
thus avoiding possible issues with colored/charged stable relics in the 
SM sector \cite{DiLuzio:2016sbl,DiLuzio:2017pfr}. 
Depending on the $\Q_k$ quantum numbers, 
this would change in turn the value of 
the electromagnetic-to-QCD anomaly ratio of the PQ current, usually denoted as 
$E/N$,  
which enters the axion-photon coupling. 

\subsubsection{Peccei-Quinn quality}
 The  threat posed on traditional QCD axion models by quantum  non-perturbative gravitational corrections~\cite{Holman:1992us,Kamionkowski:1992mf, Barr:1992qq, Ghigna:1992iv, Georgi:1981pu, Giddings:1988cx,Coleman:1988tj,Gilbert:1989nq, Rey:1989mg,Alvey:2020nyh} may also affect the models discussed here, as  $f_a$  is not very far from the Planck scale.  
These contributions are usually parametrized via effective operators, 
suppressed by powers of the Planck mass, 
that could  explicitly violate the PQ symmetry and thus spoil the solution to the strong CP problem~\cite{Holman:1992us,Kamionkowski:1992mf,Barr:1992qq,Ghigna:1992iv}.\footnote{UV 
sources of PQ breaking can be avoided 
in some invisible axion constructions within a variety of extra assumptions or 
frameworks~\cite{Randall:1992ut,Dobrescu:1996jp,Butter:2005wr,Redi:2016esr,Fukuda:2017ylt,DiLuzio:2017tjx,Fukuda:2018oco,Ibe:2018hir,Lillard:2018fdt,Ardu:2020qmo,DiLuzio:2020qio,Bonnefoy:2020llz,Bonnefoy:2018ibr}, 
or be arguably negligible 
under certain conditions~\cite{Alonso:2017avz}. 
%It is also possible to avoid the dangerous terms in ``heavy axion'' models~\cite{Rubakov:1997vp,Fukuda:2015ana,Berezhiani:2000gh,Hsu:2004mf,Hook:2014cda,Chiang:2016eav,Dimopoulos:2016lvn,Gherghetta:2016fhp,Kobakhidze:2016rwh,Agrawal:2017ksf,Agrawal:2017evu,Gaillard:2018xgk}, as their $f_a$ scale can be very low, e.g.~not far from the TeV range.
} 
  
\begin{figure}[ht]
\centering
\includegraphics[width=0.7\textwidth]{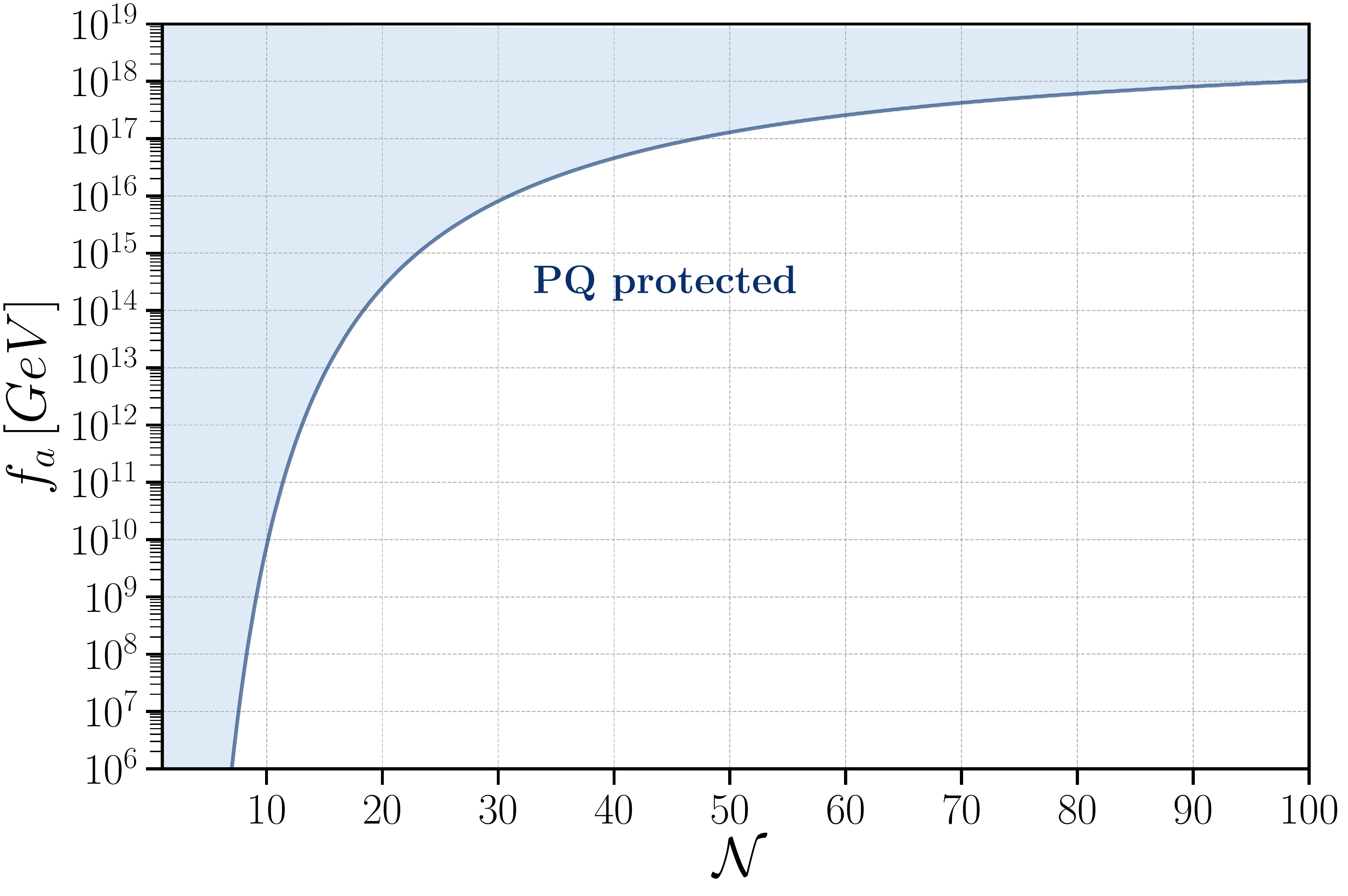} 
\caption{\small Parameter space in the $\{\N,f_a\}$ plane that is free from the PQ quality problem, within the KSVZ-like UV completion of the reduced-mass $Z_\N$ axion, for the 
PQ-breaking 
parameter values indicated in the text. The PQ protected region has a sizable overlap with the regions of parameter space where the $Z_\N$ axion can account for the total DM relic density, see Ref.~\cite{ZNDMpaper}.}
\label{fig:ZNquality}       
\end{figure}

In the context of the KSVZ $Z_\N$ axion model above, the  
exponentially small axion mass could seem to worsen this threat, increasing the sensitivity to explicit PQ-breaking effective operators.
Interestingly, promoting the {\it in built} $Z_\N$ symmetry to a gauge symmetry 
leads to an accidental $\U(1)_{\rm PQ}$ invariance, that for large $\N$ is efficiently protected from those extra 
sources of explicit breaking. 
Indeed, the lowest-dimensional PQ-violating operator 
in the scalar potential compatible with the 
$Z_\N$ symmetry is $\S^\N$, leading to an explicitly PQ-breaking contribution to the potential of the form
\beq 
V_{\rm PQ-break.} 
= c \frac{\S^\N}{M_{\rm Pl}^{\N-4}} + \text{h.c.} 
\supset \frac{\abs{c}}{2^{\N/2 -1}} \frac{f_a^\N}{M_{\rm Pl}^{\N-4}} 
\cos\( \N \theta_a + \delta \) \, ,   
\eeq
where $M_{\rm Pl} = 1.22 \times 10^{19}$ GeV is the Planck mass and  $c$ is a dimensionless coefficient with phase
$\delta \equiv \text{Arg} \ c$. 
Considering now $V_ \N(\theta_a) + V_{\rm PQ-break.}$, 
expanding for small 
$\theta_a$ the axion potential $V_ \N(\theta_a) \approx V_\N(0) + \frac{1}{2} m^2_a f^2_a \theta_a^2$,
and solving the tadpole equation, 
the induced effective $\theta$ parameter in the SM sector reads 
\begin{align}
\label{eq:thetaeff}
\vev{\theta_a} &\simeq 
\frac{\abs{c} \N f_a^\N M_{\rm Pl}^4 \sin\delta}{2^{\N/2-1} m^2_a f_a^2 M_{\rm Pl}^\N 
- \abs{c} \N^2 f_a^\N M_{\rm Pl}^4 \cos\delta} \nonumber \\ 
&\simeq
2 \sqrt{\pi} \abs{c}  \sin\delta  \sqrt{\frac{1+z}{1-z}} \frac{M_{\rm Pl}^4}{m_{\pi}^{2} f_{\pi}^{2}} \frac{1}{\sqrt{\N}}\left(\frac{f_a}{\sqrt{2}zM_{\rm Pl}}\right)^\N
\, ,
\end{align}
where 
 $m^2_a$ from \eq{Eq: hyper axion mass} has been used, and
in the last step we neglected the second term in the denominator 
 in the first line of \eq{eq:thetaeff}:  this is always justified in the $\vev{\theta_a} \lesssim 10^{-10}$ regime. 

In summary, unlike the customary ad-hoc $Z_{\N}$ protection mechanism for 
the standard KSVZ axion, in the $Z_{\N}$ axion scenario under discussion the discrete symmetry is already present by construction.  Note that
 the scaling with $\N$ is slightly different as compared to
the standard KSVZ axion,  
due to the enhancement factor $1/z^\N$. But eventually the $\( f_a / M_{\rm Pl} \)^\N$ 
suppression dominates and provides an efficient protection mechanism, 
even though the axion mass is exponentially suppressed. 
For the sake of an estimate,  \fig{fig:ZNquality}  shows
the regions in the $\{\N,f_a\}$ plane that saturate the nEDM bound for $\abs{c} =1$ and $\sin \delta = 1$.

\subsection{Composite $Z_\N$ axion} 
\label{sec:UVcompl composite}

It is also possible to construct a UV completion of the $Z_\N$ scenario which corresponds to a dynamical (composite) axion {\it \`a la} Kim-Choi \cite{Kim:1984pt,Choi:1985cb}, without extending its exotic fermionic content.  In the original version of that model, the SM fields are not charged under the PQ symmetry while two exotic {\it massless} quarks, $\psi$ and $\chi$, transform under an extra confining ``axi-color'' group $\SU(\widetilde N)_a$ and one of them, $\psi$, is also a triplet of QCD. Upon confinement of the  axi-color group 
 at a large scale $\Lambda_{a} \sim f_{a} \gg \Lambda_{\mathrm{QCD}}$, pseudo-Goldstone bosons composed of the exotic quarks emerge. All but one of them are coloured under QCD and become safely heavy. The light remaining one is the composite axion, whose mass obeys the usual formula for QCD axions Eq.~(\ref{invisiblesaxion}). 

We implement the Kim-Choi idea in the framework of our $Z_\N$ framework without increasing the number of massless exotic fermions representations. 
 The fermion $\psi$ is simply extended  to be now a triplet under {\it all} $\text{QCD}_k$ mirror sectors, see \cref{Tab:composite fermions}. The axion field will thus be unique and will couple to all anomalous terms. 
\begin{table}[h]
\begin{align}
\begin{array}{c|c|c|c|c|c|c} 
            & \SU(\widetilde N)_a  & \SU(3)_{c, \, 0}&\dots   & \SU(3)_{c, \, k}    & \dots  & \SU(3)_{c, \, \N} \\
\hline \psi & \Box      & \mathbf{3}  &\dots   &\mathbf{3}    & \dots  & \mathbf{3}  \\
\chi        & \Box      & \mathbf{1}  &\dots   & \mathbf{1}   & \dots  & \mathbf{1}  
\end{array}
\end{align}
\caption{\small Exotic fermionic sector of the $Z_\N$ composite axion model.}
\label{Tab:composite fermions}
\end{table}

Upon $\SU(\widetilde N)_a$ confinement at  
the large scale of order $ f_{a} $, the QCD$_k$ couplings $\alpha^k_{s}$ 
can be neglected, 
and therefore a large global flavor symmetry arises in the exotic fermionic sector:  $\SU(3^\N+1)_{L} \times \SU(3^\N+1)_{R} \times \U(1)_{V}$.\footnote{The $\U(1)_A$ of the exotic sector is explicitly broken by the $\SU(\widetilde N_a)$ anomaly.} This symmetry is spontaneously broken down to $\SU(3^\N+1)_{L+R} \times \U(1)_{V}$ by the exotic fermion condensates. Among the resulting
Goldstone bosons, the QCD$_k$ singlet corresponds to  the composite axion. 
Its associated PQ current reads (with $f_{\mathrm{PQ}} \equiv \widetilde N f_a$)
\begin{align}
j_{\mathrm{PQ}}^{\mu}=\bar{\psi} \gamma^{\mu} \gamma^{5} \psi-3^\N \bar{\chi} \gamma^{\mu} \gamma^{5} \chi \equiv f_{\mathrm{PQ}}\partial^{\mu} a\,,
\end{align} 
which corresponds to the only element of the Cartan sub-algebra of $\SU(3^\N+1)$ that has a vanishing anomaly coefficient with respect to $\SU(\widetilde N)_a$, but a non-vanishing one with respect to all the QCD$_k$ gauge groups.

Without further elements the model would be viable, but all mirror worlds would have the same $\theta$-parameter: a heavier than usual axion would result.  A simple $Z_\N$ implementation which leads instead to relatively shifted
potentials, and thus to a reduced axion mass, is to have a relative phase between the argument of the determinant of the quark mass matrix of the mirror worlds, 
\begin{align}
\arg\left(\det  \left(Y_u\,Y_d\right)\right)_{k+1} = \arg\left(\det  \left(Y_u\,Y_d\right)\right)_k+\frac{2 \pi }{\N}\,,
\label{Eq. shift theta param composite}
\end{align}
where $Y_u$ and $Y_d$ denote the Yukawa matrices for the up and down quark sectors, respectively. 
One of the many possible $Z_\N$ charge assignments for the quarks  that 
yield Eq.~(\ref{Eq. shift theta param composite}) is that  in which only the right-handed up quarks would transform as\footnote{Note that a factor of $1/3$ in the phase takes into account that there are 3 fermion families.}
\begin{align}
	Z_\N: U_R^k &\to e^{i {2\pi}/(3\N)}U_R^{k+1}\, , 
\label{Eq:Kimchoifertransf}
\end{align}
corresponding to a Yukawa quark Lagrangian of the form
\begin{align}
\mathcal{L}_Y = - \sum_{k=0}^{\N-1} \big\{e^{i {2\pi k}/(3\N)}\bar{Q}_{L} Y_{u} \widetilde{H} U_{R}+\bar{Q}_{L} Y_{d} H D_{R}\big\}_k + \text{h.c.} \,.
\end{align}
The resulting low-energy axion effective field theory is then the desired one as in \cref{Eq: Lagrangian ZN}.

In this $Z_\N$ composite axion model only the exotic fermions are charged under the PQ symmetry, while  the $Z_\N$ charges are carried solely by SM quarks. This means that the $Z_\N$ and PQ symmetries are not directly linked.  As a consequence, gauging the in-built $Z_\N$ symmetry would not soften the PQ quality problem,
 contrary to the case of the KSVZ $Z_\N$ axion model discussed earlier above.  Our $Z_\N$ composite axion model  is then subject to the usual PQ quality threat. Standard softening solutions often applied to composite axion models could be explored, e.g.~those based on a chiral gauging of the global symmetry of the coset space or on introducing a moose structure~\cite{Randall:1992ut,Redi:2016esr,Ibe:2018hir,Gavela:2018paw}.

\subsection{Ultra-light QCD axions}
\label{sec:ULQCDA}

The term ultra-light axions usually refers to  
the mass range $m_a \in \[ 10^{-33}, 10^{-10}\]$ eV 
(with the extrema of the interval corresponding respectively to 
an axion Compton wavelength of the size of the Hubble horizon 
and to the Schwarzschild radius of a stellar mass black hole). 
 As a theoretical motivation for ultra-light axions, 
the so-called string 
Axiverse \cite{Arvanitaki:2009fg} 
is often invoked, 
according to which a plenitude of ultra-light axions 
populating mass regions down to 
the Hubble scale $10^{-33}$ eV 
is a generic prediction of String Theory, although 
without a direct reference to the solution of the strong CP problem.\footnote{See e.g.~Ref.~\cite{Brzeminski:2020uhm} for an ultra-light scalar field whose mass is protected by a discrete $Z_N$ symmetry but does not solve the strong CP problem.}  
On the other hand, according to the usual QCD mass vs.~$f_a$  relation, Eq.~(\ref{invisiblesaxion}), 
axion masses below the peV correspond to 
axion decay constants larger than the Planck mass,  
and hence they are 
never entertained within canonical QCD axion models. 
The $Z_\N$ axion framework 
discussed in the present work 
allows in contrast to populate the sub-peV axion mass region 
while keeping sub-Planckian 
axion decay constants, with the advantage  of providing as well a direct solution to the strong CP problem.
 As shown in Sec.~\ref{subsec:finite_density_constraints}, 
 the tantalizing prospects for testing the $Z_\N$ scenario, through observational data on very dense stellar objects and gravitational waves, can sweep through the discovery region of the ultra-light axion range.

\subsection{A heavier-than-QCD axion}

A remark is in order regarding the  $Z_\N$ charge of the axion in the different sectors. 
  If the   implementation  of the $Z_\N$ symmetry would be such that the $\N$ world replicas are degenerate but the axion field is exactly the same in all of them,  that is, if \eqs{mirror-charges}{axion-detuned-charge} 
  were replaced by 
\begin{align}
	Z_\N:\quad \text{SM}_{k} &\longrightarrow \text{SM}_{k+1\,(\text{mod} \,\N)} \label{mirror-charges-again}\\
	     a &\longrightarrow a \,,
	     \label{axion-heavier-charge}
\end{align}
 the potentials of the different mirror worlds would not be relatively shifted but exactly superpose.  The axion would then be a factor $\sqrt \N$ {\it heavier} than the usual QCD axion in Eq.~(\ref{invisiblesaxion}). This scenario was proposed in Ref.~\cite{Giannotti:2005eb} 
 for a $Z_2$ symmetry with just one mirror world degenerate with the SM, but its generalization to $\N$ copies is trivial.  Such a {\it heavier-than-QCD}  axion solution is viable, and it would transform the ALP arena to the {\it right} of the canonical QCD axion band 
into solutions to the SM strong CP problem. 
  The axion $Z_\N$  charge assignment explored throughout this work, Eq.~(\ref{axion-detuned-charge}), results instead  in {\it lighter-than-QCD}  axions, that is, solutions located to the left of the QCD axion band. Note that this option 
 induces a comparatively  much larger impact: a natural {\it exponential  suppression} of the axion mass $\propto z^\N$ as the byproduct of the cancellations between the mirror potentials, \cref{Eq: hyper axion mass}, instead of the mild $\sqrt \N$ enhancement just discussed.  
 
 All in all, to explore the right-hand side region of the QCD axion band for solutions to the strong CP problem, other heavy axion scenarios proposed in the literature seem more efficient and appealing (e.g.~those with mirror worlds much heavier than the SM, or scenarios with novel confining scales much larger than $\Lambda_{\rm QCD}$, as mentioned in \sect{sec:intro}).

\begin{figure}[!h]
\centering
\includegraphics[width=0.98\textwidth]{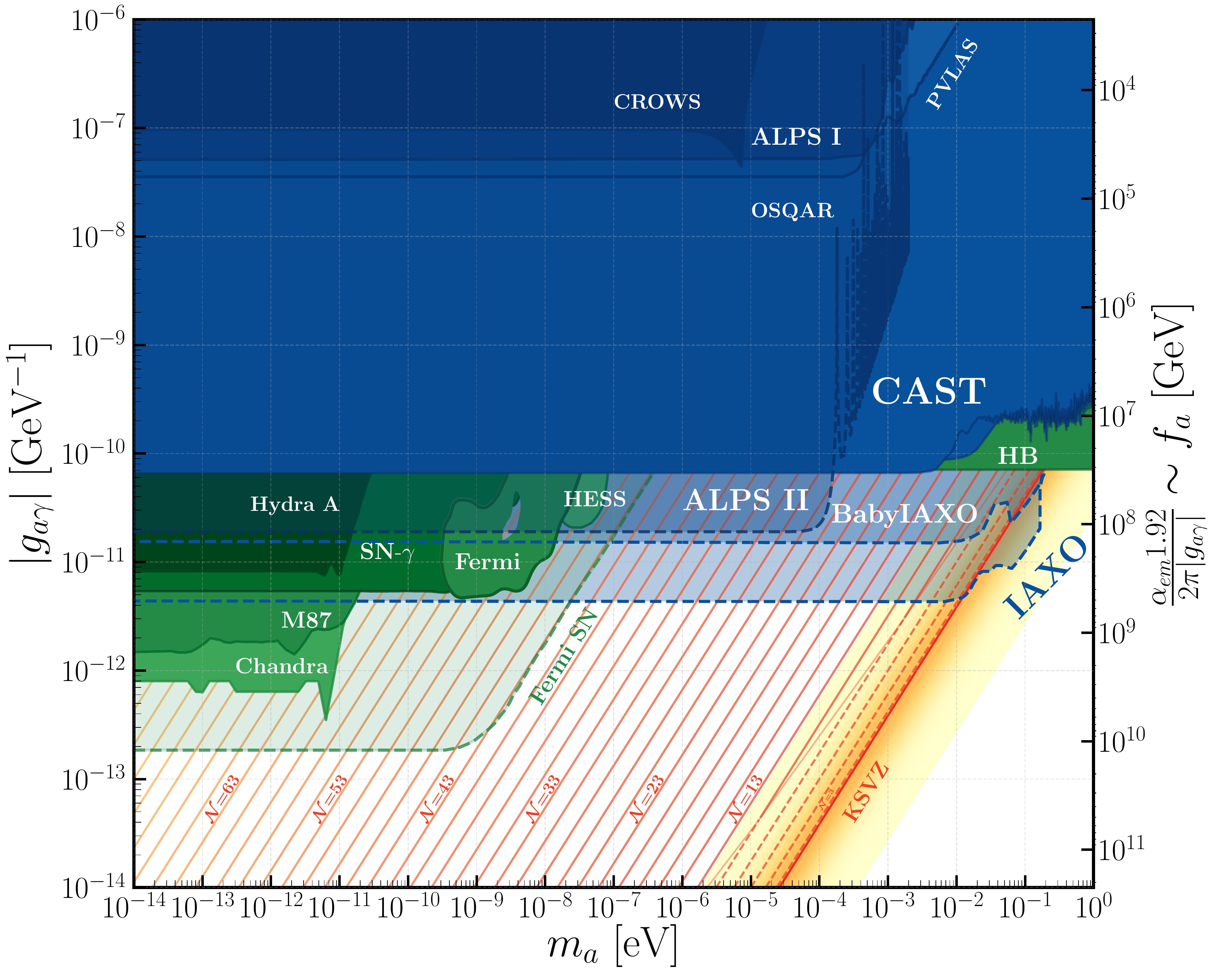} 
\caption{\small  Limits on the axion-photon coupling as a function of the axion mass. Laboratory constrains~\cite{1004.1313,arXiv:hep-ex/0702006,1705.02290,1310.8098,1212.4633,1506.08082,1510.08052,Bahre:2013ywa} and astrophysical bounds~\cite{1304.0989,1703.07354,1907.05475,2006.06722,1609.02350,1603.06978,1311.3148,1406.6053,1410.3747,Galan:2020edq,Abeln:2020ywv} are shown in {\color{blueEDM} \bf blue} and {\color{greenAstro} \bf green}, respectively. Projected sensitivities appear in translucent colors  delimited by dashed lines.  The  {\color{OrangeC2} \bf orange}  oblique lines represent the theoretical prediction for the $Z_\N$ axion photon couplings assuming $E/N=0$ for different (odd)  number of worlds $\N$. These lines are solid for the regions of the parameter space in which the KSVZ UV completion of the $Z_\N$ axion is free from PQ quality problem and dashed otherwise. 
The secondary vertical axis shows the corresponding axion decay constant $f_a$ 
if  $E/N=0$ is assumed. 
Supplementary constraints in case the axion is assumed to account for DM can be found in Ref.~\cite{ZNDMpaper}.
Axion limits adapted from Ref.~\cite{ciaran_o_hare_2020_3932430}.}
\label{fig:Photon coupling no DM}       
\end{figure}

\section{Experimental probes of down-tuned axions}
The $Z_\N$ axion with  reduced mass can provide a solution to the SM strong CP problem, independently of whether it accounts or not for the DM content of the Universe. It is hence 
interesting to get a perspective on the experimental panorama that {\it does not} rely on the supplementary assumption that the axion may be the DM particle: all experimental bounds and prospects below will 
 be independent of that hypothesis. 
On the other hand, Ref.~\cite{ZNDMpaper} will focus on experimental probes that do rely on it.

\subsection{Axion coupling to photons}

From an experimental point of view, 
a highly relevant axion coupling 
is that to photons, defined via the Lagrangian term $\delta \mathcal{L} = \frac{1}{4} g_{a\gamma} a F \tilde F$ as \cite{Georgi:1986df,diCortona:2015ldu} 
\beq\label{agammagamma_coupling}
g_{a\gamma} = \frac{\alpha}{2\pi f_a} (E/N - 1.92(4)) \, ,
\eeq
where $E$ and $N$ denote model-dependent anomalous electromagnetic and strong contributions, respectively. 
\fig{fig:Photon coupling no DM} shows the parameter space 
of the reference $Z_\N$ axion model (with $E/N=0$) 
in the coupling vs.~mass plane. Predictions 
for the axion photon coupling are obtained 
by rescaling the $Z_\N$ axion mass in \eq{Eq: hyper axion mass} 
for different values of $\N$. 
Present axion limits and projected sensitivities are displayed 
as filled and transparent areas, respectively.

 The yellow band depicts  
 the canonical QCD axion solution, which obeys the well-known relation in Eq.~(\ref{invisiblesaxion}). The oblique lines indicate instead the $Z_\N$ lighter axion solutions to the strong CP problem, 
 as a function of the number of mirror worlds $\N$, see Eq.~(\ref{Eq: hyper axion mass}). Note that {\it the overall effect of  a reduced mass axion is simply a shift towards the left of the parameter space}: each of those oblique lines can be considered to be the center of a displaced yellow 
band. 
It is particularly enticing that experiments set {\it a priori} to only hunt for ALPs may in fact be targeting solutions to the strong CP problem. 

\subsection{Finite density constraints on $f_a$}

\label{subsec:finite_density_constraints}
\begin{figure}[ht]
\centering
\includegraphics[width=0.95\textwidth]{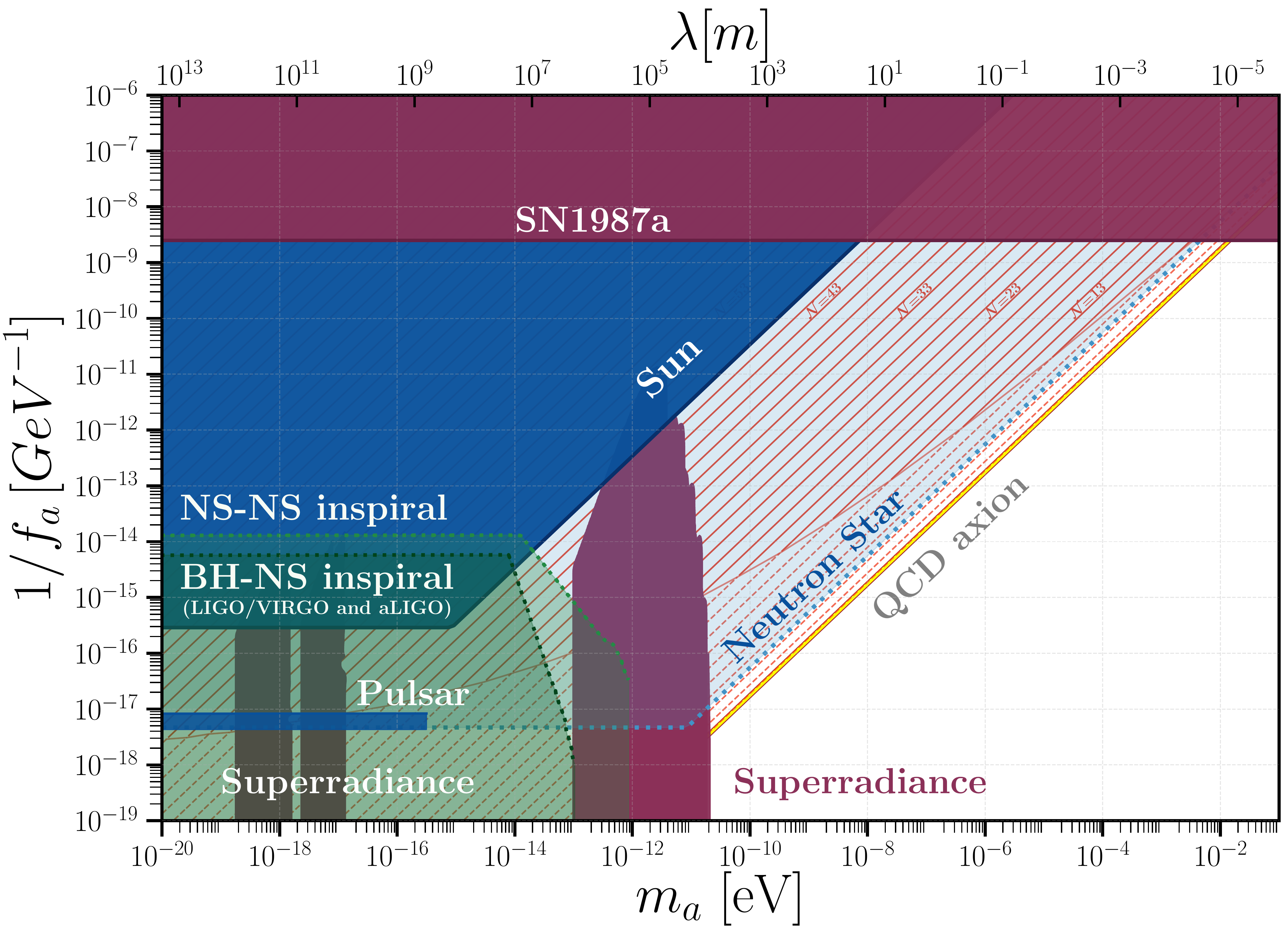} 
\caption{\small Model-independent constraints on the axion scale $f_a$ versus axion mass, from astrophysical data. Regions presently excluded are depicted in solid colors, while the translucent  regions circumscribed by dotted lines are projections. The {\color{colZN} \bf orange} oblique lines indicate the theoretical prediction of the reduced-mass $Z_\N$ QCD axion scenario,  as a function of $\N$: they are solid where the KSVZ $Z_\N$ axion is free from the PQ quality problem, and dashed otherwise. Additional constraints which apply  if the axion is assumed to account as well for DM are discussed in Ref.~\cite{ZNDMpaper}.}
\label{fig:FiniteDensity}       
\end{figure}

This subsection summarizes the model-independent constraints on $f_a$ for the  $Z_\N$ scenario under discussion.  The result of the analysis is illustrated in \cref{fig:FiniteDensity}. 
Interestingly, apart from the usual 
constraints stemming from the SN1987A~\cite{Raffelt:2006cw} and black hole superradiance measurements~\cite{Arvanitaki:2014wva,Arvanitaki:2010sy,Stott:2020gjj,Mehta:2020kwu} (depicted in purple),
 novel bounds apply to the exceptionally light $Z_\N$ axion due to finite density effects. Indeed, it has been recently pointed out in Refs.~\cite{Hook:2017psm,Huang:2018pbu} that finite density media may have a strong impact on the physics of very light axions or ALPs. In those media, the minimum of the total potential may be shifted to $\pi$. This has a number of phenomenological consequences that span from the modification of the nuclear processes in stellar objects due to $\theta \sim \mathcal{O} (1)$, to modifications in the orbital decay of binary systems (and subsequently in the emitted gravitational waves).  

For the scenario considered here, the important point is that a background made only of ordinary matter 
breaks the $Z_\N$ symmetry. This hampers 
 the symmetry-induced cancellations in the potential which led to a reduced-mass axion in vacuum: the effective axion mass will be larger within a dense medium.

We will first elaborate on the $Z_\N$ axion potential in a nuclear medium. Following Refs.~\cite{Hook:2017psm,Balkin:2020dsr}, one can compute the finite density effects on the axion potential by considering the quark condensates in a medium made of non-relativistic neutrons and protons~\cite{Cohen:1991nk,Gasser:1990ap}. Applying the Hellmann-Feynman theorem, the quark condensate at a finite density $n_N$ of a given nucleon $N$ can be expressed as 
\begin{align}
 \langle \bar q q\rangle_{n_N}= \langle \bar q q\rangle_{0}\left(1-\frac{\sigma_Nn_N}{m_{\pi}^{2} f_{\pi}^{2}}\right) \, , 
 \label{Eq: finite density condensate}
 \end{align} 
 where $\langle \bar q q\rangle_{0}=\frac{1}{2}\left( \vev{\bar u u}  + \vev{\bar d d}\right)\equiv -\Sigma_0$ is the quark condensate in vacuum --see Eq.~(\ref{condensate})-- and  $\sigma_N$ is defined by 
 \begin{align}
 \sigma_N\equiv m_q \frac{\partial M_N}{\partial m_q}\,,
 \end{align}
 \begin{figure}[ht]
\centering
\includegraphics[width=0.9\textwidth]{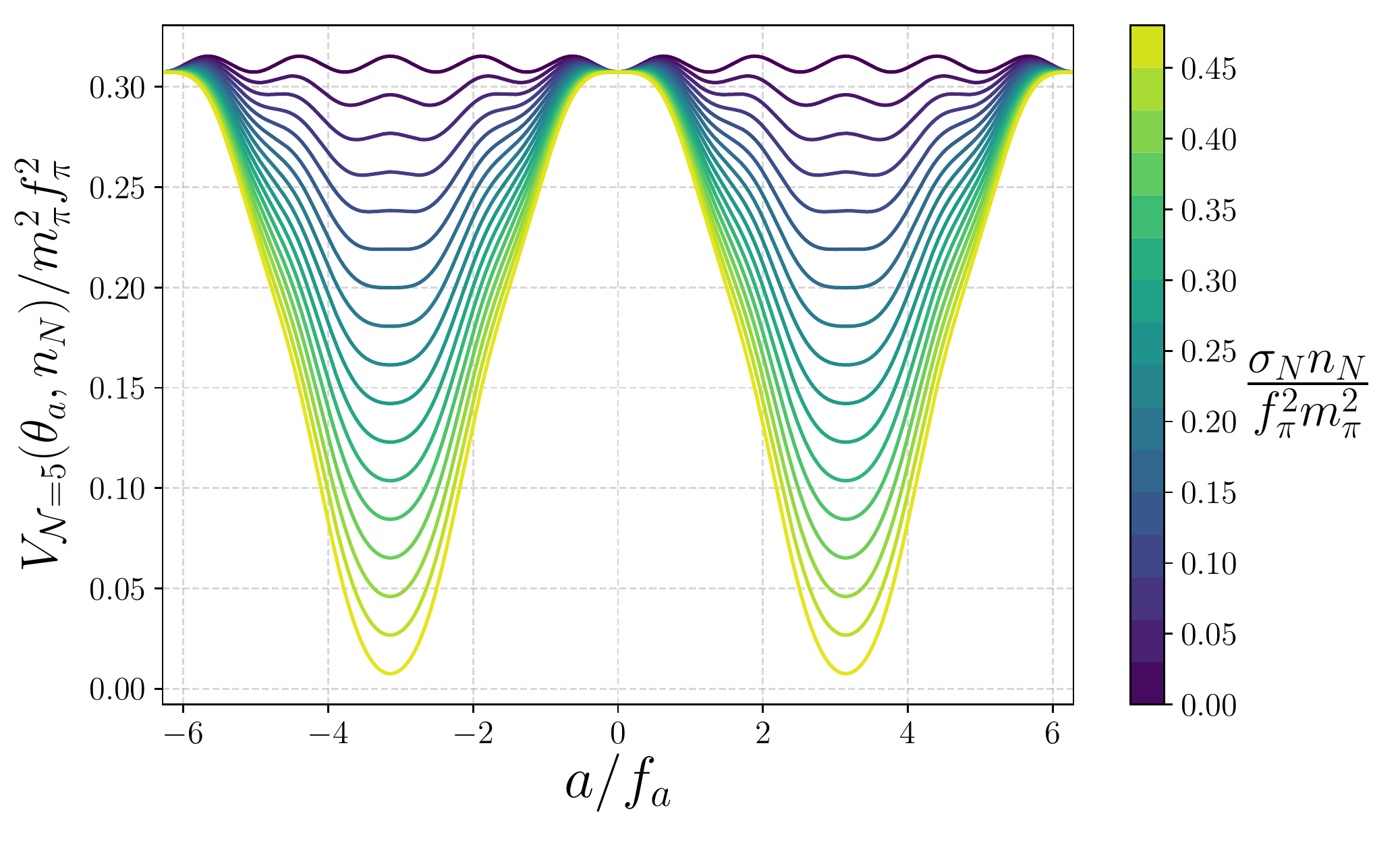} 
\caption{\small Example of the in-medium potential dependence as a function of the nuclear density for $\N=5$. For large densities (light green) the total potential develops a minimum in $\theta_a\sim\pi$.}
\label{fig:Z5 Finite density}       
\end{figure}
where $m_q\equiv \frac{1}{2}(m_u+m_d)$ and $M_N$ is the mass of the  nucleon $N$.
Because the $Z_\N$ potential is proportional to the quark condensate, see Eq.~(\ref{eq:VNpot}), we can simply obtain the potential within a SM nuclear medium $V^{f.d.}_{\mathcal{N}}\left(\theta_{a},n_N\right)$ 
 by weighting the SM (i.e.~$k=0$) contribution in the vacuum potential by the factor in \cref{Eq: finite density condensate}, that is, 
\begin{align}
\label{Eq:Finite density ZN potential}
V^{f.d.}_{\mathcal{N}}\left(\theta_{a},n_N\right) &\simeq 
 \left(1-\frac{\sigma_Nn_N}{m_{\pi}^{2} f_{\pi}^{2}}\right) V(\theta_a)+\sum_{k=1}^{\N-1}V(\theta_a+2\pi k/\N)\\
&=  -\frac{\sigma_Nn_N}{m_{\pi}^{2} f_{\pi}^{2}} V(\theta_a) + \sum_{k=0}^{\N-1}V(\theta_a+2\pi k/\N) \xrightarrow{\N\gg 1\,\,} -\frac{\sigma_Nn_N}{m_{\pi}^{2} f_{\pi}^{2}} V(\theta_a)\,. \nonumber
\end{align}
In the last step of these expressions  the large $\N$ limit has been taken, which allowed 
us to neglect the  term corresponding to the exponentially reduced axion potential in vacuum (see Eq.~\ref{Eq: fourier potential large N hyper}). This shows that, if the nucleon density is large enough,  
the  $Z_\N$ asymmetric background spoils the cancellations among the mirror world contributions to the potential, in such a way that the total potential in matter is proportional to {\it minus} the SM one in vacuum $V(\theta_a)$. Therefore, the minimum of the potential is located at  $\theta_a=\pi$.
More precisely,  
 \begin{align}
V^{f.d.}_{\mathcal{N}}\left(\theta_{a},n_N\right)  
  \xrightarrow{\N\gg 1\,\,} \frac{m_{\pi}^{2} f_{\pi}^{2}}{1+z}\bigg[\frac{\sigma_Nn_N}{m_{\pi}^{2} f_{\pi}^{2}} \sqrt{1 + z^2 + 2 z \cos{\( \theta_a \)}} \, -\frac{\mathcal{N}^{-1 / 2}z^{\mathcal{N}}}{\sqrt{\pi}} \sqrt{1-z^2}  \cos \left(\mathcal{N} \theta_{a}\right)\bigg]\,,
\end{align}
 which requires
 \beq
 \frac{\sigma_Nn_N}{m_{\pi}^{2} f_{\pi}^{2}}\gg z^\N
 \eeq
for the minimum to sit at $\theta_a=\pi$. This is illustrated in \cref{fig:Z5 Finite density}.

A large value of the $\theta$ parameter 
inside dense stellar objects is rich in physical consequences, which translates into strong constraints for the $Z_\N$ scenario.
As it was pointed out in Ref.~\cite{Hook:2017psm}, $\theta \sim \mathcal{O}(1)$ 
 inside the solar core is excluded due to the increased proton-neutron mass difference (which would prohibit the neutrino line corresponding to the Be$^7$-Li$^7$ mass difference  observed by Borexino~\cite{Bellini:2013lnn}). Similarly, for $\theta\sim \pi$ in nearby neutron stars (NS), $\mathrm{Co}^{56}$ would be lighter than $\mathrm{Fe}^{56}$~\cite{Ubaldi:2008nf,Lee:2020tmi} and therefore $\mathrm{Fe}^{56}$ could have been depleted due to its $\beta$-decay to $\mathrm{Co}^{56}$. The presence of iron in the surface of neutron stars and its implications in terms of the allowed $\theta$ values could be explored through dedicated X-ray measurements~\cite{Mukai:2017qww}. The corresponding current and projected constraints that were derived in Ref.~\cite{Hook:2017psm} (within the simplifying assumption $z=1$) are translated here to the $Z_\N$ scenario and further generalized for any $z$.

A conservative criterion consistent with  $\theta=\pi$ inside the medium is to  impose that the axion mass at $\theta_a=0$ becomes tachyonic, i.e.~$-m_T^2>0$ where $m_T^2$ is defined by 
\begin{align}
-m_T^2\equiv\frac{d^2V^{f.d.}_{\mathcal{N}}}{d^2a}\Bigg|_{\theta_a=0}=
\frac{m_\pi^2 f_\pi^2}{f_a^2}\left[\frac{1}{\sqrt{\pi}} \,\sqrt{\frac{1-z}{1+z}} \,\N^{3/2} \,z^\N - \frac{\sigma_Nn_N}{m_{\pi}^{2} f_{\pi}^{2}}  \frac{z}{(1+z)^2} \right]\,.
\label{Eq: tachyonic}
\end{align}
Requiring this quantity to be positive, it directly follows a limit on the number of  worlds 
allowed by the stellar bounds above:
\begin{equation}
\N\lesssim 47\,, 
\end{equation}
where the most recent estimation of $\sigma_N$ has been used.\footnote{
 We employ here  $\sigma_N\simeq59 \, \mathrm{MeV}$ which is in agreement with recent determinations based on Roy-Steiner equations $\sigma_N=59.1(3.5) \, \mathrm{MeV}$ \cite{Hoferichter:2015dsa} and ChPT estimates $\sigma_N=59(7) \, \mathrm{MeV}$ \cite{Alarcon:2011zs}. } 
 This bound does not apply for the whole range of $f_a$, though, because the argument only makes physical sense as long as the reduced Compton wavelength of the axion is smaller than the stellar object, $r_{core}\gtrsim 1/m_a^{f.d.}$, where $m_a^{f.d.}\sim1/f_a$ is the effective axion mass in the medium,
\begin{align}
 \big(m_{a}^{f.d.}\big)^2=\frac{d^2V^{f.d.}_{\mathcal{N}}}{d^2a}\Bigg|_{\theta_a=\pi}
=\frac{m_\pi^2 f_\pi^2}{f_a^2}\left[ \frac{\sigma_Nn_N}{m_{\pi}^{2} f_{\pi}^{2}}  \frac{z}{1-z^2} - \frac{1}{\sqrt{\pi}} \,\sqrt{\frac{1-z}{1+z}} \,\N^{3/2} \,z^\N  \right]\,.
\end{align}
For the case of the sun, $r_{core}\sim 139.000$ km implies $f_a \lesssim 2.4 \times 10^{15}\text{ GeV}$ for the observational constraints to apply. Finally, the area in parameter space excluded is depicted in dark blue in \cref{fig:FiniteDensity}.
Analogously, the future sensitivity prospects 
from neutron star data are depicted in shaded blue.\footnote{Our results are analogous to those in Eq.~(1.7) of Ref.~\cite{Hook:2017psm}, with their generic parameter $\epsilon$  identified 
as $
\epsilon = {m_a^2}/{m_a^2(\N=1)} \simeq {\pi}^{-1/2}\sqrt{1-z^2} (1+z)\, \N^{3/2} z^{\N-1}$. 
Note that the location of the QCD axion line, as well as our projected exclusion regions for neutron stars and gravitational waves, are shifted  towards the left by a factor of five with respect to those in  Refs.~\cite{Hook:2017psm,Huang:2018pbu}.}

Even stronger bounds may be established by relaxing the requirement stemming from \cref{Eq: tachyonic}. Indeed, as it can be seen in \cref{fig:FiniteDensity}, long before the mass in $\theta_a=0$ becomes tachyonic, the absolute minimum of the potential corresponds to $\theta\sim \mathcal{O}(1)$. Therefore one could constrain larger masses or smaller $\N$ values in the $Z_\N$ scenario than those obtained above. This would require, however, 
a dedicated analysis to ensure that the axion field would fall into the absolute minimum, so as to overcome the potential barrier; this development lies beyond the scope of the present work.

The fact that the position of the minimum of the axion potential depends on the nuclear density of the medium not only modifies the effective $\theta$-parameter inside stellar objects but  may also source a long-range force between them. This has been studied in Refs.~\cite{Hook:2017psm,Huang:2018pbu}. This new long-range force sourced by the axion can be constrained by the measurement of double pulsar  or neutron star (NS) - pulsar binaries~\cite{Taylor:1982zz,Burgay:2003jj,Will:2001mx}. Moreover, the existence of these axionic long-range forces would modify the gravitational wave signal emitted by NS-NS mergers or black hole-NS mergers which will be probed in the future by LIGO/VIRGO and Advanced LIGO \cite{Huang:2018pbu,PhysRevLett.119.161101}. The projected constraints from Ref.~\cite{Huang:2018pbu} are show in green in \cref{fig:FiniteDensity}.  It is striking that the whole ultra-light DM region, included the so-called ``fuzzy dark matter'' region ($m_a\sim10^{-22}$ eV)~\cite{Hui:2016ltb}, will be within observational reach in the next decades, 
for a wide range of $\N$ values.

\section{Conclusions}
\label{sec:conclusions}

An axion which solves the strong CP problem may be much lighter than the canonical QCD axion, down to the range of ultra-light axions, provided Nature has a $Z_\N$ symmetry implemented via $\N$ degenerate world copies, one of which is our SM. The axion field realizes the symmetry non-linearly, which leads to exponential cancellations among the contributions from each mirror copy to the total axion potential.    
For large $\N$, we have shown that the total axion potential is given by a single cosine and we  
determined analytically the --exponentially suppressed-- dependence of the axion mass on the number of mirror worlds, using the properties of hypergeometric functions and the Fourier expansion. In practice, the formula in \eq{Eq: hyper axion mass}
gives an excellent approximation even down to $\N=3$. We have also improved the holomorphicity bounds previously obtained. 

We compared next the predictions of the theory with present and future data from experiments which do not rely on the additional assumption that an axion abundance may explain the DM content of the Universe.    It is particularly enticing that experiments set {\it a priori} to hunt only for ALPs may in fact be targeting solutions to the strong CP problem.  For instance,  
ALPS II is shown to be able to probe the $Z_\N$ scenario here discussed down to 
$\N \sim 25$ for a large enough axion-photon coupling, while IAXO and BabyIAXO may test the whole $\N$ landscape for  values of that coupling even smaller, see Fig.~\ref{fig:Photon coupling no DM}. In turn, Fermi SN data  can only reach  $\N \gtrsim 43$ but are sensitive to smaller  values of the coupling. 

Highly dense stellar bodies allow one to set even stronger bounds in wide regions of the parameter space. These exciting limits have an added value: they avoid model-dependent assumptions about the  axion couplings to SM particles, because they rely exclusively on the anomalous axion-gluon interaction needed to solve to the strong CP problem. A dense medium of ordinary matter is a background that breaks the $Z_\N$ symmetry. This hampers the symmetry-induced cancellations in the total axion potential:
 the axion becomes heavier inside dense media {\it and} the minimum of the potential is located at $\theta_a=\pi$.  From present solar data we obtain  the bound $\N\lesssim 47$ provided $f_a\lesssim 2.4\times 10^{15}\text{ GeV}$, while larger $\N$ values are allowed for smaller $f_a$. Furthermore, we showed that projected neutron star and pulsar data should allow to test the scenario down to $\N \sim 9$ --and possibly even below-- for the whole range of $f_a$, see \cref{fig:FiniteDensity}. Furthermore, gravitational wave data from NS-NS and BH-NS mergers by LIGO/VIRGO and Advanced LIGO will allow to probe all values of $\N$ for the remaining $f_a$ range, up to the Planck scale and including the ultra-light axion mass range. 

These analytical and phenomenological results have been derived within the model-independent framework of effective couplings. Nevertheless, for the sake of illustration, we have developed two examples of UV completed models. One is a $Z_\N$ KSVZ model, which is shown to enjoy an improved PQ quality behaviour: its $Z_\N$ and PQ symmetries are linked and in consequence gauging $Z_\N$ alleviates much the PQ quality problem. The other UV completion 
considered in this paper  is a $Z_\N$ version of the composite axion {\it \`a la} Kim-Choi. 
While this model is viable, its PQ quality is not improved with respect to the usual situation, because its $Z_\N$ and PQ symmetries are independent.

This work is intended to be a proof-of-concept that a much-lighter-than usual axion is a viable solution to the strong CP problem, with spectacular prospects of being tested in near future data. It also pinpoints that  experiments searching for generic ALPs  have in fact  discovery potential to solve the strong CP problem. 

The down-tuned axion considered here could also explain the DM content of the Universe in certain regions of the parameter space. The impact of such a light axion on the cosmological history is significant and it will be discussed in a separate paper~\cite{ZNDMpaper}.

\begin{small}

\section*{Acknowledgments}
We thank Gonzalo Alonso-\' Alvarez, Victor Enguita, Mary K.~Gaillard, Yann Gouttenoire, Benjamin Grinstein, Lam Hui, David B.~Kaplan, D. Marsh, V. Mehta, Philip S{\o}rensen and  Neal Weiner for illuminating discussions.  
M.B.G.~and P.Q.~are indebted for hospitality to the Theory Department of Columbia University in New York, where the initial stage of their work took place.
The work of L.D.L.~is supported by the Marie Sk\l{}odowska-Curie Individual Fellowship grant AXIONRUSH (GA 840791). L.D.L., P.Q.~and A.R.~acknowledge support by 
the Deutsche Forschungsgemeinschaft under Germany's Excellence Strategy 
- EXC 2121 Quantum Universe - 390833306.
M.B.G.~acknowledges support  from the ``Spanish Agencia Estatal de Investigaci\'on'' (AEI) and the EU ``Fondo Europeo de Desarrollo Regional'' (FEDER) through the projects FPA2016-78645-P and PID2019-108892RB-I00/AEI/10.13039/501100011033.
M.B.G. and P.~Q. acknowledge support from the European Union's Horizon 2020 research and innovation programme under the Marie Sklodowska-Curie grant agreements 690575  (RISE InvisiblesPlus) and  674896 (ITN ELUSIVES), as well as from  
 the Spanish Research Agency (Agencia Estatal de Investigaci\'on) through the grant IFT Centro de Excelencia Severo Ochoa SEV-2016-0597. This project has received funding/support from the European Union's Horizon 2020 research and innovation programme under the Marie Sklodowska-Curie grant agreement No 860881-HIDDeN.

\appendix

%\newpage

\section{Holomorphicity properties of $Z_\N$ axion potential}

\label{app:math}

In order to determine the parameter $b$ in \cref{Eq: theorem En}, which  controls the exponential suppression of the axion mass,  it is necessary to study the region in the complex plane where the extension of the potential $V(w)$ is holomorphic. As  the plots in \cref{fig:holomV} illustrate, both the potential and its second derivative have branch cuts starting in $w_{cut}=\pi \pm i \log z$. However, the second derivative $V''(w)$ diverges at the branch point and thus $b$ cannot be extended all the way to $\log z$. In order to optimize the bound on the axion mass we allow $b$ to depart from $\log z$, $b=| \log z+ \Delta b\,|$. Taking into account that $V''(w)$ for small $\Delta b$ can be approximated by
\begin{align}
V''\big(\pi + i(\log z+ \Delta b)\big)\simeq - m_{\pi}^{2} f_{\pi}^{2} \sqrt{\frac{1-z}{1+z}}\left[\frac{1}{4} \frac{1}{\left(\Delta b\right)^{3/2}} + \mathcal{O}\big(\Delta b^{-1/2}\big)\right]\,, 
\label{Eq: app V'' delta b}
\end{align}
  the procedure amounts to minimize the function $B(\Delta b) $ that determines the bound $\left|E_{\mathcal{N}}(V)\right| \leq B(\Delta b) $ (see \cref{Eq: theorem En}, 
\begin{align}
B(\Delta b)\equiv  \frac{4 \pi M(\Delta b)}{e^{\N |\log z+ \Delta b\,|}-1}=\pi m_{\pi}^{2} f_{\pi}^{2} \sqrt{\frac{1-z}{1+z}} \,\frac{1}{\left(\Delta b\right)^{3/2}}\frac{1}{e^{\N| \log z -\Delta b \,|}-1}\,. 
\end{align}
The requirement $\frac{d B (\Delta b)}{d\left(\Delta b\right)}=0$ shows that the bound is optimized for  
\begin{align}
\Delta b= \frac{3}{2}\frac{1}{\N}\,, 
\end{align}
where the factor $3/2$ comes form the order of the divergence in \cref{Eq: app V'' delta b}.
\begin{figure}[ht]
\centering
\includegraphics[width=0.49\textwidth]{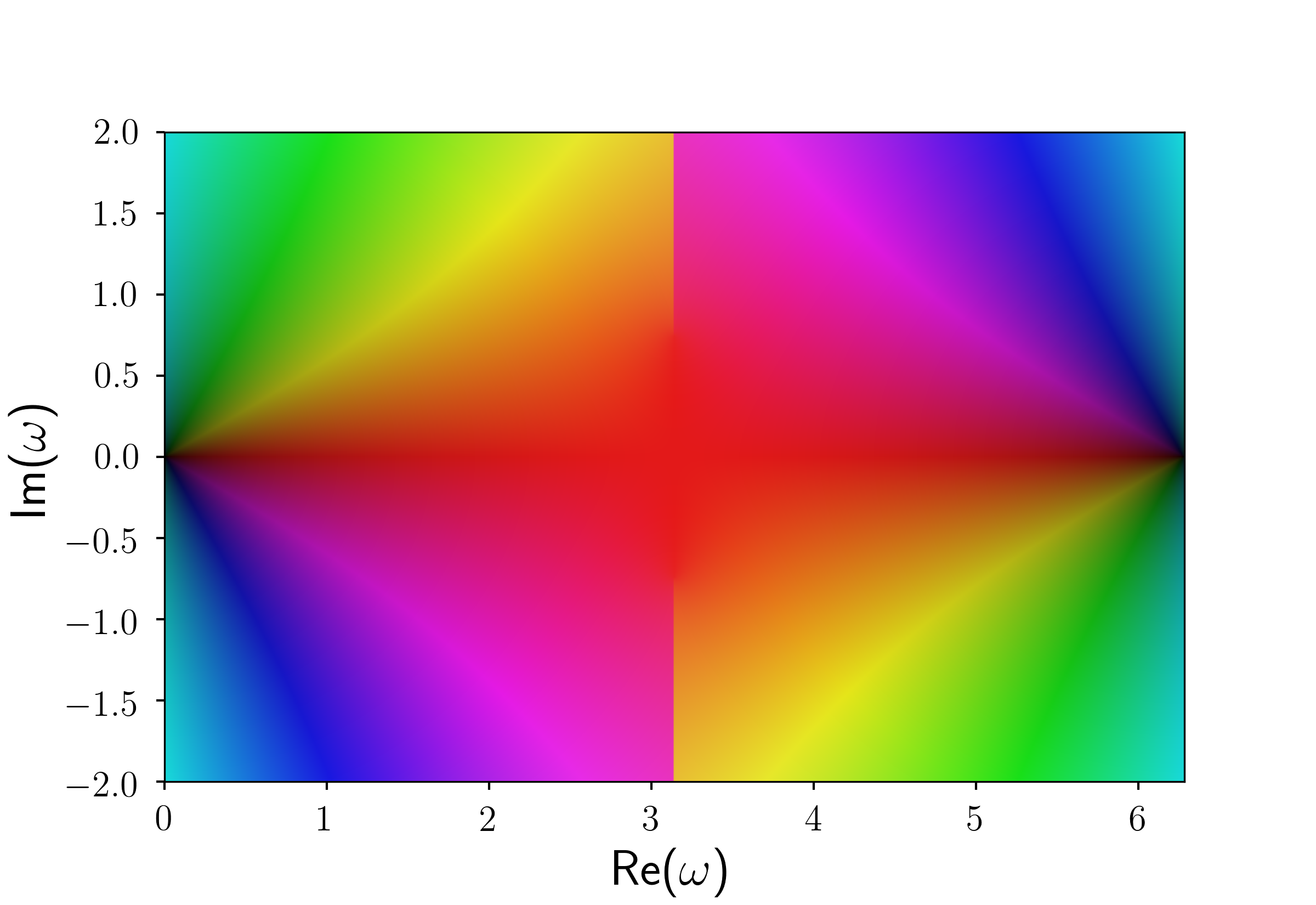} \includegraphics[width=0.49\textwidth]{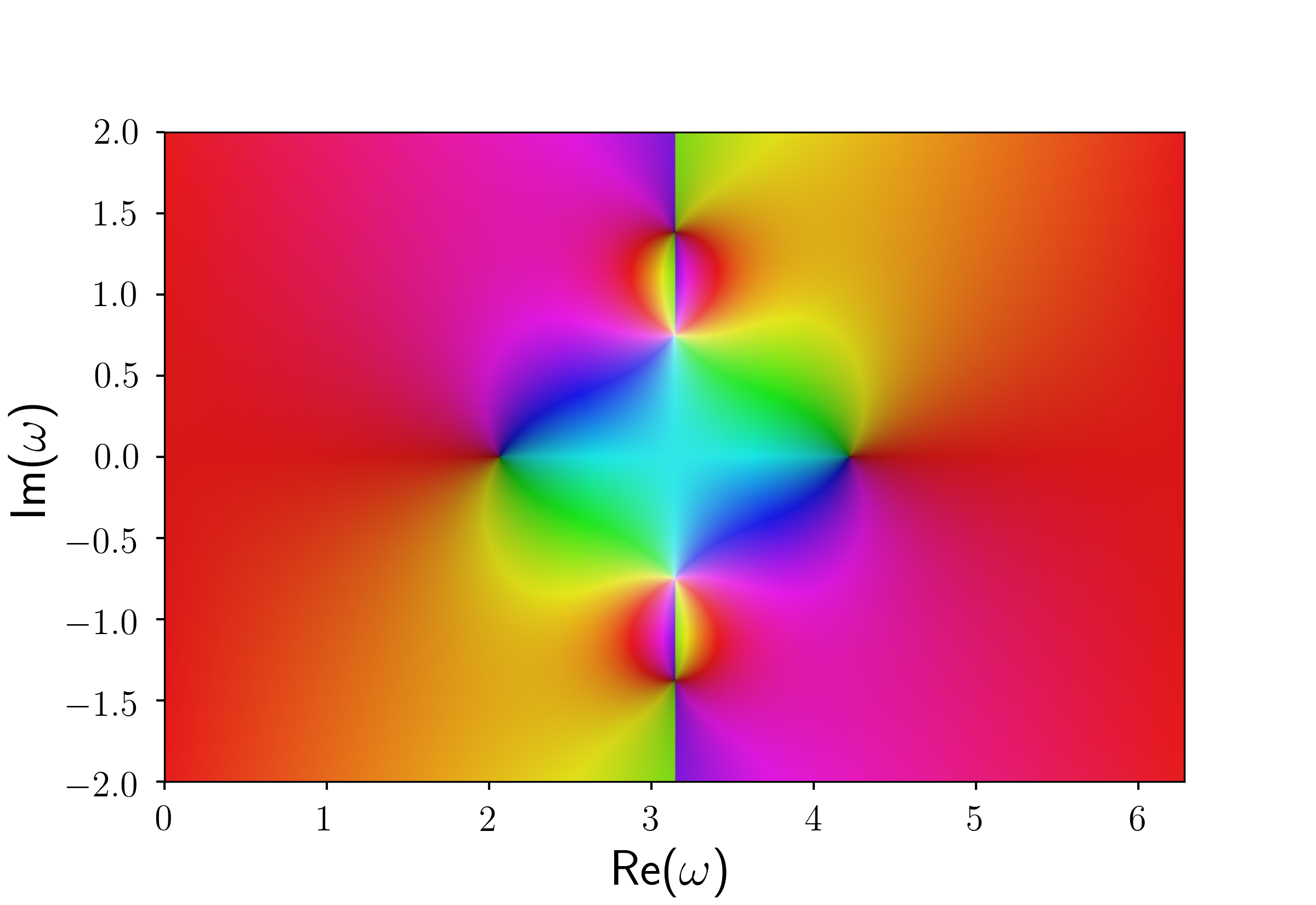} 
\caption{\small Representation of the complex functions $V(w)$ (left) and $V''(w)$ (right). Colors represent the phase of the corresponding  complex function and the brightness represents the modulus. The singularities can be clearly spotted: branch cuts starting from $w_{cut}=\pi \pm i \log z\,$ in both functions and divergences in those same points for $V''(w)$.}
\label{fig:holomV}       
\end{figure}

%\newpage
\section{Fourier series of the $Z_\N$ axion potential}
\label{App: fourier}

We show here that the coefficients of the Fourier series of any $Z_\N$ symmetric potential, such as the $Z_\N$ axion potential in \cref{eq:VNpot},
 vanish unless the corresponding Fourier mode is a multiple of $\N$. Moreover it will be shown that, when the potential is expressed as 
 \begin{equation}
  V_{\mathcal{N}}\left(\theta_{a}\right)= \sum_{k=0}^{\N-1} V\left(\theta_a+\frac{2 \pi k}{\mathcal{N}}\right)\,,
  \label{Eq: General ZN potential}
\end{equation}
 the non-vanishing coefficients of the Fourier series can be expressed in terms of the Fourier transformation of a single  term in the sum \cref{Eq: General ZN potential}.

 Let us denote by $ \hat V_{\N}(n)$ the coefficients of the Fourier series of the total potential, 
   \begin{equation}
  V_{\mathcal{N}}\left(\theta_{a}\right)\equiv
\sum_{n=-\infty}^{\infty} e^{in\theta_a} \hat V_{\N}(n) \, ,
\label{VN}
 \end{equation}
 and by  $\hat{V}_{{2 \pi k}/{\mathcal{N}}}(n)$  the coefficients of the Fourier series of each of the terms in the sum in Eq.~(\ref{Eq: General ZN potential}),
  \begin{align}
 V\left(\theta_a+\frac{2 \pi k}{\mathcal{N}}\right)\equiv\sum_{n=-\infty}^{\infty} \hat{V}_{{2 \pi k}/{\mathcal{N}}}(n) e^{in\theta_a}\,. 
 \label{Eq:Definition fourier-1}
\end{align}
We will stick to the notation that omits the subindex  for the first world ($k=0$), $ \hat{V}_0(n)\equiv \hat{V}(n)$,
\begin{align}
 V(\theta_a)=\sum_{n=-\infty}^{\infty} \hat{V}(n) e^{in\theta_a} \quad\text{ with }\quad  \hat{V}(n)=\frac{1}{2\pi}\int_0^{2\pi} V(x) e^{-inx}dx\,.
 \label{Eq:Definition fourier}
\end{align}
Each term in the sum in Eq.~(\ref{Eq: General ZN potential}) and \cref{Eq:Definition fourier-1} corresponds to the function in the first term but with its argument shifted by $\frac{2 \pi k}{\mathcal{N}}$.  The shift property  of the Fourier series  relates the Fourier  coefficients of the shifted functions $ \hat{V}_{{2 \pi k}/{\mathcal{N}}}(n)$
to that of the original function,
\begin{align}
 \hat{V}_{{2 \pi k}/{\mathcal{N}}}(n)= e^{in\frac{2 \pi k}{\mathcal{N}}}\hat V(n) \, .
\end{align}
  Substituting this expression in  Eq.~(\ref{Eq:Definition fourier-1}), and inserting the latter in Eq.~(\ref{Eq: General ZN potential}), it follows that the total potential can be written as 
  \begin{equation}
V_{\mathcal{N}}\left(\theta_{a}\right)
= \sum_{k=0}^{\N-1} \sum_{n=-\infty}^{\infty} \hat V_{{2 \pi k}/{\mathcal{N}}}(n) e^{in\theta_a}=
  \sum_{n=-\infty}^{\infty} \hat V(n) e^{in\theta_a} \, \sum_{k=0}^{\N-1}  e^{in\frac{2 \pi k}{\mathcal{N}}} \, .
 \end{equation}
  Comparing this expression with Eq.~(\ref{VN}), it follows that the coefficients of the Fourier series for the total potential are given by
  \begin{align}
\hat V_{\N}(n)= \hat V(n) \sum_{k=0}^{\N-1}  e^{in\frac{2 \pi k}{\mathcal{N}}} \, .
\end{align}
Interestingly, these coefficients vanish unless $n$ is a multiple of $\N$
\begin{align}
\text{If } n\,(\text{mod} \, \N)\neq 0 \,
&\implies\, \sum_{k=0}^{\N-1}e^{in\frac{2 \pi k}{\mathcal{N}}}=0 \,
\implies \hat{V}_{\mathcal{N}}(n)=0 \, , \\
\text{If } n\,(\text{mod} \, \N)= 0 \,
&\implies \, \sum_{k=0}^{\N-1}e^{in\frac{2 \pi k}{\mathcal{N}}}=\N 
\implies \hat V_{\N}(n)= \N \hat V(n) \, .
\end{align}
To sum up, the Fourier series of the total potential $V_{\N}(\theta)$ can be easily obtained in terms of the Fourier series of a single term $V(\theta)$ 
  and it only receives contributions from the modes that are multiples of $\N$. In our case of interest the potential is real and even, this translates into
\begin{align}
V_{\mathcal{N}}\left(\theta_{a}\right)=2 \N 
	\sum_{t=1}^{\infty} \hat V(t\N)  \cos (t\,\N\theta_a) \, ,
\label{Eq: fourier potential general}
\end{align}
where the factor of two comes from the negative modes and the constant term (i.e.~$\theta_a$-independent) has been obviated.

\section{Analytical axion mass dependence from hypergeometric functions}
\label{App: Hypergeometric complete mass dependence}
We show here that the Fourier series coefficients of the single world axion potential in \cref{Eq: fourier single world def},
\begin{align}
\hat V(n)=-  \frac{m_{\pi}^{2} f_{\pi}^{2}}{1+z}\int_0^{2\pi} \cos(n t)\sqrt{1+z^{2}+2 z \cos \left(t\right)}dt\,,
\label{Eq: fourier single world defApp}
\end{align}
 can be written for large Fourier modes, $n\gg 1$, as a simple analytical formula that exponentially decays with $n$. Moreover, by applying the result in \cref{App: fourier}, it will be shown that the potential for the $Z_\N$ axion approaches a single cosine and a simple formula for the  $Z_\N$ axion mass follows.

Let us start by relating the Fourier series decomposition of the single world potential in \cref{Eq: fourier single world defApp} with the Gauss hypergeometric functions (see for example Eq.~(9.112) in Ref.~\cite{GRADSHTEYN1980904}),
\begin{align}
\,_2F_{1}\left(\begin{array}{c}
p ,\,n+p\\
 n+1
\end{array} \bigg|\,w^2\right)=
\frac{w^{-n}}{2 \pi} \frac{\Gamma(p) n !}{\Gamma(p+n)} \int_{0}^{2 \pi} \frac{\cos (n t)\, d t}{\left(1-2 w \cos t+w^{2}\right)^{p}}  \,.
\end{align}
Via the identification $w=-z$ and $p=-1/2$, $\hat V(n)$ can be written as
\begin{align}
\hat V(n)=(-1)^{n+1}  \frac{m_{\pi}^{2} f_{\pi}^{2}}{1+z}\, z^{n} \,\frac{\Gamma(n-1/2)}{\Gamma(-1/2) \,n !}\, 
\,_2F_{1}\left(\begin{array}{c}
-1/2,\, n-1/2\\
 n+1
\end{array} \bigg|z^2\right) \, . 
\label{Eq: fourier hypergeome}
\end{align}
For convenience, the hypergeometric function can be also expressed as (see Eq.~(9.131) from Ref.~\cite{GRADSHTEYN1980904})
\begin{align}
\,_2F_{1}\left(\begin{array}{c}
\alpha,\, \beta\\
\gamma
\end{array} \bigg|\,w\right)
=(1-w)^{-\alpha}\,\,_2F_{1}\left(\begin{array}{c}
\alpha,\, \gamma-\beta\\
\gamma
\end{array} \bigg|\,\frac{w}{w-1}\right) \, , 
\end{align}
so that
\begin{align}
\,_2F_{1}\left(\begin{array}{c}
-1/2,\, n-1/2\\
n+1
\end{array} \bigg|\,z^2\right)
=\left(1-z^2\right)^{1/2} 
\,_2F_{1}\left(\begin{array}{c}
-1/2,\, 3/2\\
n+1
\end{array} \bigg|\,\frac{z^2}{z^2-1}\right) \, .
\end{align}

The relation  in Eq.~(\ref{Eq: fourier hypergeome}) is exact. 
However, only the modes $n$ which are multiples of $\N$ contribute to the potential (see \cref{App: fourier}), and therefore it is pertinent to focus on the large $n$ limit. 
While the limit of the Gauss hypergeometric function  when one or more of its parameters become large is difficult to compute in general,
some asymptotic expansions of the hypergeometric function are known in the literature. In particular, following Ref.~\cite{Cvitkovi2017,article},
\begin{align}
\lim_{\gamma\to\infty}\,_2F_{1}\left(\begin{array}{c}
\alpha,\, \beta\\
\gamma
\end{array} \bigg|\,w\right)
=1+\frac{\alpha\beta}{\gamma}w +\mathcal{O}\left((w/\gamma)^2\right) \, . 
\end{align}
In turn, the prefactor in \cref{Eq: fourier hypergeome} simplifies in the large $n$ limit to 
\begin{align}.
\lim_{n\to\infty} \frac{\Gamma(n-1/2)}{\Gamma(-1/2) \,n !}= -\frac{1}{2\sqrt{\pi}}n^{-3/2}\,.
\end{align}
Putting all this together, it follows that, in the large $n$ limit, the coefficient of the Fourier series for a single world is given  by
\begin{align}
\hat V(n)=\,(-1)^{n}\, \frac{m_{\pi}^{2} f_{\pi}^{2}}{2\sqrt{\pi}} \, \sqrt{\frac{1-z}{1+z}}n^{-3/2}\, z^{n} \, ,
\end{align}
which in turn implies in this limit that the total $Z_\N$ potential in \cref{Eq: fourier potential general} can be written as
\begin{align}
V_{\mathcal{N}}\left(\theta_{a}\right)= \N 
	\sum_{t=1}^{\infty} \,(-1)^{t\,\N}\, \frac{m_{\pi}^{2} f_{\pi}^{2}}{2\sqrt{\pi}} \, \sqrt{\frac{1-z}{1+z}}\left(t\,\N\right)^{-3/2}\, z^{t\,\N}  \cos (t\,\N\theta_a) \, .
\label{Eq: fourier potential generalApp}
\end{align}
This expression allows us to understand several properties of the total potential. Firstly,   it can be shown now that 
the total potential approaches a single cosine in the large $\N$, since all the other modes are then exponentially suppressed with respect to the first one,  
\begin{align}
\lim_{\N\to\infty}\bigg|\frac{\hat V_\N(t\,\N)}{\hat V_\N(\N)}\bigg|
=\lim_{\N\to\infty}\bigg|\frac{\hat V(t\,\N)}{\hat V(\N)}\bigg|=t^{-3/2}z^{(t-1)\N} \longrightarrow 0 \, ,
\end{align}
and thus the potential reads 
\begin{align}
V_{\mathcal{N}}\left(\theta_{a}\right) \xrightarrow{\N\gg 1\,\,} \frac{m_{\pi}^{2} f_{\pi}^{2}}{\sqrt{\pi}} \, \sqrt{\frac{1-z}{1+z}}\N^{-1/2}\,.
   \,(-1)^{\N}\,  z^{\N}  \cos (\N\theta_a)\,,
\label{Eq: fourier potential large N hyperApp}
\end{align}
Secondly, we can also obtain an analytical expression for the axion mass that confirms the dependence obtained from the holomorphicity arguments in \cref{Subsec: Holomorphicity bounds}, and completes the expresion with the correct prefactor,
\begin{equation}
  \label{Eq: hyper axion massApp}
  m_a^2f^2_a  \simeq  \frac{m_{\pi}^{2} f_{\pi}^{2}}{\sqrt{\pi}} \,\sqrt{\frac{1-z}{1+z}} \,\N^{3/2} \,z^\N\,. 
\end{equation} 
Finally, it is now trivial to show that  the potential in the large $\N$ limit  has $\N$ minima (maxima)  located at $ \theta_a =\{\pm {2\pi \ell/\N}\}$  for $\ell=0,1,\dots,\frac{\N-1}{2}\,,$ for odd (even) $\N$. 

The results above assumed the large $\N$ limit. However, note that the conclusion about the location of the extrema is true for any $\N$. This can be easily seeing after obtaining the {\it exact} Fourier expansion of the $Z_\N$ axion potential in \cref{eq:VNpot}, which reads,
\begin{align}
V_{\mathcal{N}}\left(\theta_{a}\right)= -m_{\pi}^{2} f_{\pi}^{2} \,\mathcal{N} \sum_{t=1}^{\infty} (-1)^{t\N+1}  \sum_{\ell=t\N}^{\infty} \frac{(2 \ell) !(2 \ell) !}{2^{4 \ell-1}(2 \ell-1)(\ell !)^{2}(\ell-t\N) !(\ell+t\N) !} \beta^{\ell} \cos \left(t \,\mathcal{N} \theta_{a}\right)\,.
\end{align}
For even $\N$, it trivially follows that $\theta_a=0$ is a maximum, as all factors 
  in this expression are positive except for the factor  $(1-2\ell)<0$. For odd $\N$ instead the  $ (-1)^{t\,\N}$ factor fluctuates in sign, but  the first term ($t=1$) is positive and dominates the expansion (e.g.~it is exponentially larger in magnitude  than the $t=2$ term which is negative). The periodicity of the potential extends these conclusions to the location of all extrema.

\bibliographystyle{utphys.bst}
\bibliography{bibliography}

\end{small}

\end{document}